\documentclass[twocolumn,twocolappendix, resetfootnote]{aastex701}
\usepackage{xcolor}
\usepackage{graphicx}
\usepackage{dcolumn}
\usepackage{bm}
\usepackage{float}
\usepackage{hyperref}
\usepackage[caption=false]{subfig}
\usepackage{mathtools}
\usepackage{mathrsfs} 
\usepackage[noabbrev, capitalise]{cleveref}

\graphicspath{{./}{figures/}}
\begin{document}

\title{Gravitational Waves from Strongly Magnetized Eccentric Neutron Star Binaries}

\author[0000-0002-6602-3913]{R. Prasad}
\affiliation{International Centre for Theoretical Sciences,
    Tata Institute of Fundamental Research, Bangalore 560089, India}
\email[show]{prasad.r@icts.res.in}  

\author[0000-0003-3895-7994]{Anushka Doke}
\affiliation{International Centre for Theoretical Sciences,
    Tata Institute of Fundamental Research, Bangalore 560089, India}
\affiliation{ Department of Physics, University of Massachusetts, Dartmouth, Massachusetts 02747, USA}
\affiliation{Center for Scientific Computing and Data Science Research, 
University of Massachusetts, Dartmouth, Massachusetts 02747, USA}
\email{adoke@umass.edu}

\author[0000-0001-5523-4603]{Prayush Kumar}
\affiliation{International Centre for Theoretical Sciences,
    Tata Institute of Fundamental Research, Bangalore 560089, India}
\email{prayush@icts.res.in}

\begin{abstract}
We explore the dynamics of neutron star binaries that approach their final inspiral stages with residual eccentricity and strong magnetic fields, features that can arise in systems formed through dynamical capture of relatively young neutron stars. Our analysis focuses on identifying magnetic-field imprints on the gravitational wave signal arising from two mechanisms: magnetic interaction between the neutron stars and electromagnetic radiation from the system’s effective dipole. Using a perturbative approach, we obtain the associated gravitational wave energy loss rate and phase evolution, and quantify detectability through cumulative dephasing, horizon distances, and Fisher-matrix analyses. While magnetic effects are intrinsically small, entering at 2 post-Newtonian (PN) order, their cumulative influence over extended inspirals will become distinguishable in future detectors owing to enhanced low-frequency sensitivity. For binaries with comparable magnetic fields, we show that $10^{14}\,\mathrm{G}$ systems will be detectable up to $\sim10$ Mpc with DECIGO and the Einstein Telescope, while $10^{15}\,\mathrm{G}$ fields will be discernible out to several hundred Mpc. For extreme fields of $10^{16}\,\mathrm{G}$, third-generation detectors could probe out to Gpc scales. These findings suggest that magnetic effects in compact binaries can indeed become observable with next-generation detectors, offering a potential probe of magnetar-level fields and binary formation pathways.
\end{abstract}



\section{Introduction}\label{section:introduction}

The detection of gravitational waves from compact binary mergers has opened a transformative observational window into the universe. So far, around 218 events have been detected during the O1, O2, O3, and O4a observing runs~\citep{abbott2019gwtc,abbott2021gwtc, abbott2023gwtc, abac2025gwtc}, spanning binary black hole (BBH), neutron star–black hole (NSBH), and binary neutron star (BNS) mergers. Observations of BBH events have enabled measurements of component masses and spins~\citep{tong2022population, nitz20234, abbott2023gwtc}, revealing a population of black holes more massive~\citep {barack2019black, fishbach2022apples, liotine2023missing} than those inferred from electromagnetic observations of low-mass X-ray binaries~\citep{bailyn1998mass, farr2011mass}. Gravitational wave detections have also confirmed the existence of NSBH systems, hybrid binaries comprising a black hole and a neutron star, through multiple merger events~\citep{abbott2021observation}. In particular, such systems have not yet been identified in electromagnetic surveys, such as in pulsar timing and X-ray observations~\citep{manchester2005australia, booth2009meerkat, nan2011five, jonker2011galactic, amiri2022overview, moon2024nustar}. These findings highlight the ability of gravitational wave astronomy to uncover compact object populations that previously eluded detection by conventional electromagnetic searches.

The two BNS mergers GW170817~\citep{abbott2017gw170817} and GW190425~\citep{abbott2020gw190425} have provided stringent constraints on the properties of neutron stars~\citep{abbott2018gw170817, abbott2019properties, shibata2019constraint, abbott2020gw190425, chatziioannou2020neutron, nathanail2021gw170817}. Although the total mass of GW170817 is consistent with that observed in 19 known galactic BNS systems, GW190425 exhibited a relatively higher total mass $\sim 3.4\, M_\odot$, prompting questions about the formation of such high-mass systems and their merger rates~\citep {foley2020updated, romero2020origin, zhu2020characterizing, korol2021can}. In addition, a few compact objects detected via gravitational waves appear to reside in the so-called mass gap~\citep{abbott2020gw190814, abac2024observation}. A striking example is the secondary of GW190814~\citep{abbott2020gw190814}, with a mass of approximately $2.6\, M_{\odot}$, which could be an unusually massive neutron star or an exceptionally light black hole~\citep{most2020lower, fattoyev2020gw190814, dexheimer2021gw190814, godzieba2021maximum}. Collectively, these events advance our understanding of compact object populations~\citep{abbott2019binary, roulet2019constraints, abbott2021population, abbott2023population}, and offer new avenues to probe gravity~\citep{yunes2013gravitational, yagi2016black, chamberlain2017theoretical, perkins2021probing, johnson2022investigating}, matter under extreme conditions~\citep{andersson2011gravitational, chatziioannou2015probing, bose2018neutron, khadkikar2025cosmic}, and the formation pathways of compact binaries~\citep{belczynski2002comprehensive,ivanova2008formation, tagawa2020formation,samsing2014formation, mapelli2022formation}. 

Gravitational wave signals from binary neutron stars encode rich imprints of their structure, tidal response, and composition~\citep{faber2002measuring, flanagan2008constraining, read2009measuring, bauswein2012measuring, chatziioannou2022uncertainty}. While magnetic fields of neutron stars are often neglected in standard waveform models, they can introduce subtle cumulative modifications to orbital dynamics and the shape of the gravitational wave signal. Studies have explored the impact of magnetic fields on the inspiral phase of compact binaries, generally finding that very strong magnetic fields $ B\geq 10^{16} \, \mathrm{G}$ produce a noticeable effect to cause template mismatches~\citep{ioka2000gravitational, giacomazzo2009can, zhu2020tidal, tang2024prospect}. However, these studies have remained largely restricted to binaries in circular orbits and short inspirals~\citep{ioka2000gravitational, tang2024prospect}. Also, the possibility of such strong fields in merging binaries was considered less plausible, since in binaries formed via isolated channels, long evolutionary timescales will allow for magnetic field decays, reducing them to levels ($\sim 10^{9–12}$ G) where their influence becomes negligible~\citep{ioka2000gravitational, giacomazzo2009can}. 

{\it In contrast, because of their much shorter formation-to-observation times, dynamically formed neutron star binaries could retain strong magnetic fields close to their merger}. This scenario has received less attention. With next-generation detectors on the horizon, magnetic effects previously beyond reach may become observable in extended and high-amplitude inspiral signals. Motivated by these considerations, this work explores the dynamics of magnetized neutron star binaries that approach their final inspiral stages with residual eccentricity and strong magnetic fields, features that may arise in systems formed outside isolated evolutionary channels.

Interestingly, clues of eccentricity have already emerged in a few of the detected gravitational wave events, although they remain inconclusive~\citep{romero2021signs,romero2022four,gupte2024evidence,planas2025eccentric}. Theoretical studies predict that a significant fraction of eccentric mergers could be detected in the future~\citep{dhurkunde2025search}. These systems are expected to form through dynamical interactions in dense environments, such as globular clusters and galactic disks around supermassive black holes, where the concentrations of compact objects are high~\citep{park2017black, mapelli2020binary, grobner2020binary}. With next-generation ground-based detectors, like the Einstein Telescope (ET) and Cosmic Explorer (CE), that offer larger horizon distances and improved sensitivity~\citep{punturo2010einstein,hall2022cosmic}, a larger population of dynamically captured binaries could be observed. These future observatories will also significantly enhance our ability to probe the properties of neutron stars. For example, radius measurements can be obtained with a remarkable accuracy of $50–100$ meters for individual events~\citep{bandopadhyay2024measuring, huxford2024accuracy}. Additionally, upcoming space-based missions such as DECIGO~\citep{kawamura2011japanese, yagi2011detector} and LISA~\citep{danzmann1996lisa} will further expand the observable population. Thus, it is imperative to systematically investigate the influence of magnetic fields in inspiralling BNS systems and assess the prospects for detecting such effects.

Although the persistence of strong magnetic fields remains uncertain, their possible survival in dynamically formed binaries has interesting consequences. In such binaries, magnetic fields may play a more substantial role during the inspiral phase than previously thought, with dynamical effects that extend beyond their association with pre-merger precursors such as fast radio bursts \citep{most2020electromagnetic,suvorov2024premerger}. The short gamma‑ray burst accompanying BNS mergers requires strong magnetic fields, existing pre‐merger or generated post‑merger. Multimessenger observations, complemented by inspiral‑phase gravitational‑wave signatures of magnetic fields, could thus also provide stronger clues on the central engine powering these bursts. Furthermore, in events where even tidal deformability cannot distinguish BHNS systems from BNS systems, magnetic fields may serve as an independent diagnostic.

Accounting for these effects is also crucial for accurate parameter inference and minimizing biases in gravitational wave analysis~\citep{littenberg2013systematic, vallisneri2013stealth, moore2014novel, gair2015quantifying, varma2017effects, cho2022systematic, wang2024anatomy}, especially as observations continue to reveal diverse binary populations with unexpected features and extreme configurations~\citep{abac2025gw231123}.

Magnetized compact binaries exhibit two primary effects: the interaction between magnetic moments and the electromagnetic emission from moving dipoles. While the former has been examined in several works~\citep{vasuth2003gravitational, bourgoin2022impact, savalle2024detection}, the latter has been largely overlooked. A recent analytic framework by~\cite{henry2024electromagnetic} incorporates both effects, but focuses on eccentric white dwarf binaries in the LISA band with aligned magnetic orientations. In contrast, binary neutron star systems, particularly those with strong fields and misaligned moments, remain less explored, despite their potential for richer dynamical behavior.

In this work, we analytically compute the orbital evolution of a compact binary system with magnetic moments undergoing eccentric inspiral, focusing on the distinct contributions of magnetic interaction, electromagnetic emission, and orientation-dependent effects to the orbital dynamics and gravitational wave signal. For simplicity, tidal interactions are neglected, and the analysis is performed in the perturbative regime, where magnetic effects enter as corrections to the leading-order dynamics. We identify the regions of the parameter space where each magnetic effect becomes dominant. We derive the total energy loss due to both gravitational wave emission and electromagnetic dipole radiation, along with the corresponding phase evolution of the gravitational wave signal. Importantly, our formulation retains arbitrary magnetic moment orientations, preserving the full angular dependence in the analytical expressions, an essential ingredient for accurately modeling generic binary configurations. This aspect is particularly relevant for dynamically formed systems, where magnetic moments are expected to be randomly aligned~\citep{stevenson2017hierarchical, farr2017distinguishing}. Our analysis also extends beyond galactic binaries to distant sources, covering both short and long-lived inspirals with frequencies between decihertz and kilohertz.

By capturing relevant magnetic field-induced dynamics in eccentric binaries, this work contributes to ongoing efforts to develop waveform templates that more faithfully reflect real astrophysical systems~\citep{favata2010gravitational, chatziioannou2017constructing,pang2018potential,brito2018black, talbot2018gravitational, wadekar2023new}. Furthermore, we identify regions of the parameter space where current and future ground-based detectors, as well as upcoming decihertz observatories, are expected to be sensitive to these effects, thereby enabling their detection and constraining the magnetic field strength. We find that magnetar-level fields, $10^{14}-10^{15}$ G, will be very much measurable with ET and DECIGO, while LIGO will only be sensitive to ultrastrong fields, $\gtrsim 10^{16}$ G, within their observational frequency band for GW170817-like distances. These thresholds are lower by a few orders of magnitude than previous estimates of discernible magnetic-field effects \citep{ioka2000gravitational}.

The outline of the paper is as follows. Section~\ref{section:mag-fields-in-NS} describes the current understanding of neutron star magnetic fields. Section~\ref{section:eq-of-motion} presents the equations of motion for binary neutron star systems with magnetic effects. Section~\ref{section:energy-loss-eqs} discusses energy emission through both gravitational and electromagnetic radiation, followed by an analysis of waveform dephasing in Section~\ref{section:dephasing}. The results and discussions are described in Section~\ref{section:results-discussion}, and the conclusions are presented in Section~\ref{section:conclusion}.

\section{Magnetic fields in neutron stars}\label{section:mag-fields-in-NS}

Neutron stars are highly magnetized compact objects. Magnetic fields of neutron stars are estimated by combining theoretical models and observational data, which suggest typical values of $10^{11–13}$ G for slow-spinning pulsars and $10^{8–9}$ G for fast-spinning pulsars~\citep{ reisenegger2001magnetic, borghese2020exploring}. A class of neutron stars, known as \textit{magnetars}, exhibits extreme magnetic fields, typically in the range of $10^{14–15}$ G~\citep{olausen2014mcgill, turolla2015magnetars, kaspi2017magnetars}. There is mounting evidence for even stronger magnetic fields~\citep{stella2005gravitational,makishima2014possible,makishima2024observational}. Constraints on magnetars as progenitors for short gamma-ray bursts allow for magnetic field strengths approaching $10^{17}$ G~\citep{quirola2024probing}. The theoretical upper bound derived using the virial theorem places the maximum magnetic field that a neutron star can support at $ B_{\text{max}} \sim 10^{18}$ G~\citep{lai1991cold, reisenegger2008neutron, chandrasekhar2013hydrodynamic}, which is two to three orders of magnitude higher than the strongest fields currently observed. This limit arises from balancing the magnetic pressure against the star's gravitational binding energy (see Appendix~\ref{app:max-mag-field-virial}). Numerical solutions of the Einstein-Maxwell equations for rotating neutron stars also support this upper bound~\citep{bocquet1995rotating, cardall2001effects}. Simulations of core-collapse supernovae, too, have demonstrated that magnetic fields exceeding $10^{15} \, \mathrm{G}$ can emerge~\citep{akiyama2003magnetorotational, raynaud2020magnetar}.

\begin{figure}[!htbp]
    \centering
    \includegraphics[scale=0.45]{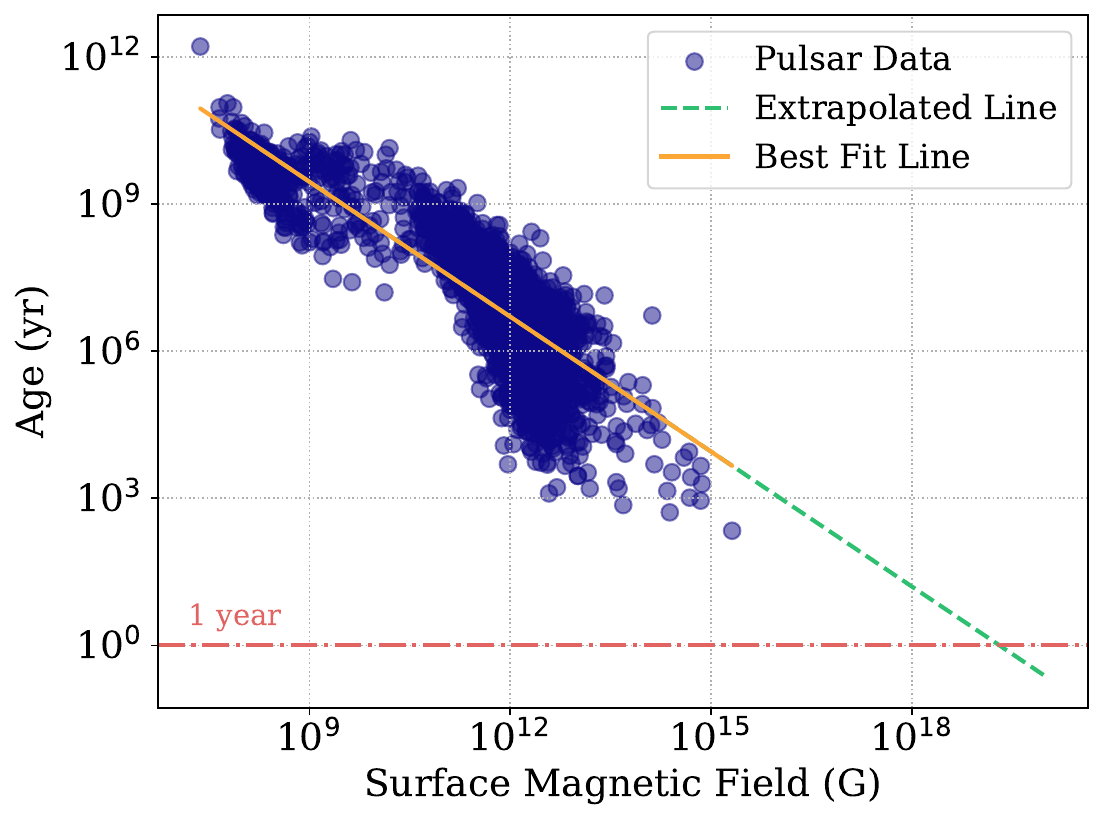}
    \caption[atnf]{A representation of the age vs magnetic field for pulsars. The orange line is the linear fit to the data, which has been extrapolated (the green dashed line). The red line denotes the one-year age line. Data used is from the Australia Telescope National Facility (ATNF) Pulsar Catalogue~\citep{manchester2005australia}.}
    \label{fig:atnf_pulsar_study_plot}
\end{figure}

To explore the observational landscape, we analyzed pulsars listed in the Australia Telescope National Facility (ATNF) Pulsar Catalogue~\citep{manchester2005australia}. A linear fit between the characteristic age and magnetic field strength of these pulsars (see~\cref{fig:atnf_pulsar_study_plot}) suggests that very young pulsars can harbour ultra-strong magnetic fields.\footnote{Characteristic age $ t_c = P / 2\dot{P}$, though an approximate indicator, is used here as a proxy for the true age. Throughout this paper, \textit{magnetic field} refers to the surface magnetic field.} Although such ultra-strong fields have not been directly observed, both observational trends and theoretical predictions support the plausibility of neutron stars possessing magnetic fields well above the typical range.

Magnetic fields in neutron stars are believed to decay over time as a result of internal processes such as Ohmic loss, Hall drift, and ambipolar diffusion~\citep{goldreich1992magnetic, bhattacharya2002evolution,reisenegger2003origin,igoshev2021evolution}. This decay takes place over timescales of $10^{6–9}$ years and is commonly modeled by~\citep{heyl1998common,malov2001magnetic}:
\begin{equation}
B(t) = B_{0} e^{-t/t_D},
\end{equation}
where $B_{0}$ is the initial magnetic field and $t_D$ is the decay timescale. The population of pulsars suggests trends consistent with magnetic field decay, such as a reduction in magnetic field strength with increasing characteristic age, but direct evidence at the level of individual sources remains lacking. Furthermore, both the rate and the extent of the magnetic field decay are highly uncertain and depend sensitively on the internal and structural properties of the neutron star. High crustal conductivity, low internal temperatures, and stable magnetic field configurations have been shown to prolong decay timescales, whereas elevated temperatures, crustal impurities, complex field geometries, and episodes of accretion are associated with more rapid dissipation~\citep{urpin1992crustal, geppert1994accretion, konenkov2000effect, pons2007evidence, pons2007magnetic, pons2009magneto}. General relativistic effects and core superfluidity may further support the long-term retention of strong magnetic fields~\citep{geppert2000magnetic, konar2002magnetic, graber2015magnetic}. Consequently, the long-term evolution of magnetic fields in neutron stars remains an important open question in astrophysics~\citep{bransgrove2018magnetic, vigano2021magneto, skiathas2024combined}. 
 
For binaries consisting of neutron stars, magnetic field decay is central in determining whether strong fields persist until the late stages of inspiral. In systems formed through isolated channels, where two main-sequence stars eventually evolve into neutron stars, the time until merger is typically very long $\sim\mathrm {Gyr}$~\citep{belczynski2018origin, zhu2020characterizing}. As a result, even if neutron stars are initially highly magnetized, by the time they reach the late inspiral phase, the magnetic field strength may decrease to around $\sim 10^{9–12}$ G. At such levels, the influence of magnetic fields on gravitational waves from inspiraling binaries becomes negligible. In binaries formed through dynamical capture channels, the inspiral time could be significantly shorter~\citep{zhu2020characterizing}. If this inspiral timescale is less than the magnetic field decay timescale $t_{D}$, the magnetic field can remain appreciably stronger at the final stages, potentially affecting both the dynamics of the system and the emitted gravitational waves. Moreover, such systems may retain measurable eccentricity~\citep{huerta2013effect,  samsing2014formation}, further distinguishing them observationally. 

Magnetic fields also influence post-merger phenomena. The collimated outflows and short gamma-ray bursts observed in BNS merger events require post-merger magnetic field strengths of at least $\sim 10^{15}$ G~\citep{kawamura2016binary, ciolfi2020key}. Such extreme fields can arise if the progenitor neutron stars were already strongly magnetized~\citep{ruiz2016binary, aguilera2024delayed} or they can be generated through efficient amplification mechanisms, such as the Kelvin-Helmholtz instability and magnetorotational turbulence, that may operate during the merger process~\citep{giacomazzo2015producing, kiuchi2015efficient, musolino2025impact, gutierrez2025magnetic}. Joint observations of gamma-ray bursts and measuring magnetic fields via gravitational waves can help shed light on the progenitor mechanism.

The presence (or absence) of ultra-strong magnetic fields in gravitational wave-emitting neutron star binaries can provide essential insights into the generation, evolution, and upper limits of neutron star magnetic fields. The gravitational-wave signal may contain distinct imprints of these magnetic fields, which could provide us with a new way to measure them. In the following section, we derive the equations governing the dynamics of magnetized neutron star binaries in eccentric orbits using a perturbative approach.


\section{Equations of Motion}\label{section:eq-of-motion}

We consider a binary neutron star system with masses $m_{i}$ and magnetic moments $\mu_{i}$. The total mass of the system and the symmetric mass ratio are given by $m=m_{1}+m_{2}$ and $\eta=m_{1}m_{2}/m^{2}$. We assume the neutron stars to be nonspinning. The Lagrangian of the system can be written as \citep{ioka2000gravitational}
\begin{equation}\label{eq:lagrangian}
    L=\frac{1}{2}\,\eta m v^{2} + \frac{\eta m^{2}}{r} + \frac{1}{r^{3}} \, [ 3 (\hat{n}.\vec{\mu}_{1}) (\hat{n}.\vec{\mu}_{2}) - (\vec{\mu}_{1}. \vec{\mu_{2}})],
\end{equation}
where $\vec{r}$ is the orbital separation between the stars, $\hat{n}=\vec{r}/r$, and $v^{2}= (\dot{r}^2 + r^{2} \dot{\phi}^2)$ is the relative velocity. From the Euler-Lagrange equations, we obtain the radial equation of motion
\begin{equation}\label{eq:eom_r}
    \eta m \ddot{r} = \eta m r \dot{\phi}^2 - \frac{\eta m^2}{r^2} - \frac{3}{r^4} \left[ 3(\hat{n} \cdot \vec{\mu}_1)(\hat{n} \cdot \vec{\mu}_2) - \vec{\mu}_1 \cdot \vec{\mu}_2 \right].
\end{equation}
For the angular coordinate, the equation of motion corresponds to the conservation of angular momentum and can be written as
\begin{equation}\label{eq:eom_phi}
\frac{d}{dt}(\eta m r^2 \dot{\phi}) = 0.
\end{equation}
To simplify the calculation, we express the time derivatives of $r$ in terms of derivatives with respect to $\phi$ using the relations:
\begin{align}
    \dot{r} &= \frac{dr}{dt} = \frac{dr}{d\phi} \frac{d\phi}{dt} = \frac{dr}{d\phi} \dot{\phi}, \label{eq:rdot} \\
    \ddot{r} &= \frac{d}{dt} \left( \frac{dr}{d\phi} \dot{\phi} \right) = \frac{d^2 r}{d\phi^2} \dot{\phi}^2 + \frac{dr}{d\phi} \ddot{\phi}. \label{eq:rddot}
\end{align} 
From~\cref{eq:eom_phi}, we obtain $\ddot{\phi} = - \dfrac{2 \dot{r} \dot{\phi}}{r}$. Substituting for $\dot{r}$, this becomes $\ddot{\phi} = -\dfrac{2 \dot{\phi}^2}{r} \dfrac{dr}{d\phi}$. Inserting this into~\cref{eq:rddot}, the second derivative of $r$ simplifies as 
\begin{align}\label{eq:rdoubledot}
\ddot{r} = \left[ \frac{d^2 r}{d\phi^2}   - \frac{2}{r} \left( \frac{dr}{d\phi} \right)^2 \right] \dot{\phi}^2.
\end{align}
Substituting~\cref{eq:rdoubledot} into the radial equation of motion,~\cref{eq:eom_r}, we obtain
\begin{align}
\eta m &\Bigg[ \left( \frac{d^2 r}{d\phi^2} \right)
- \frac{2}{r} \left( \frac{dr}{d\phi} \right)^2 \Bigg] \dot{\phi}^2 
= \eta m r \dot{\phi}^2 \nonumber \\
&- \frac{\eta m^2}{r^2} 
- \frac{3}{r^4} \left[ 3 (\hat{n} \cdot \vec{\mu}_1)(\hat{n} \cdot \vec{\mu}_2) 
- \vec{\mu}_1 \cdot \vec{\mu}_2 \right],
\end{align}
where all derivatives are expressed with respect to the angular coordinate $\phi$. We define the magnetic interaction term as $\mathcal{F}_{\mathrm{int}}(\vec{\mu}_{1}, \vec{\mu}_{2}) = \left\{ 3 (\hat{n} \cdot \vec{\mu}_{1}) (\hat{n} \cdot \vec{\mu}_{2}) - (\vec{\mu}_{1} \cdot \vec{\mu}_{2}) \right\} $. Since $(\hat{n} \cdot \vec{\mu}_{1})(\hat{n} \cdot \vec{\mu}_{2})$ varies over an orbit, we replace it with its orbit-averaged form:
\begin{align}\label{eq:orbit-averaged-expr}
 \left\langle (\hat{n} \cdot \vec{\mu}_{1})(\hat{n} \cdot \vec{\mu}_{2}) \right \rangle &= \frac{1}{2} \left\{ (\vec{\mu}_{1} \cdot \vec{\mu}_{2}) - (\hat{L} \cdot \vec{\mu}_{1})(\hat{L} \cdot \vec{\mu}_{2}) \right\}  \nonumber \\ & + \, \frac{3}{8} \, e^{2} \left[ (\vec{\mu}_{1} \cdot \hat{p})(\vec{\mu}_{2} \cdot \hat{p}) - (\vec{\mu}_{1} \cdot \hat{q})(\vec{\mu}_{2} \cdot \hat{q}) \right]
\end{align}
where $\hat{L}$ denotes the unit vector along the orbital angular momentum, and $(\hat{p},\hat{q})$ forms an orthonormal basis in the orbital plane, with $\hat{p}$ pointing along the peraistron and $\hat{q}$ perpendicular to it (see Appendix~\ref{app:orbit-averaging} for calculation). Substituting~\cref{eq:orbit-averaged-expr} in the definition of $\mathcal{F}_{\mathrm{int}}(\vec{\mu}_{1}, \vec{\mu}_{2})$, the orbit-averaged magnetic interaction becomes
\begin{align}
\bar{\mathcal{F}}_{\mathrm{int}} (\vec{\mu}_{1}, \vec{\mu}_{2}, e)  = \bar{\mathcal{F}} (\vec{\mu}_{1}, \vec{\mu}_{2}) + e^2 \bar{\mathcal{G}} (\vec{\mu}_{1}, \vec{\mu}_{2})
\end{align}
with
\begin{align}
\bar{\mathcal{F}}(\vec{\mu}_{1}, \vec{\mu}_{2}) 
&= \frac{1}{2}\Big[(\vec{\mu}_{1}\!\cdot\!\vec{\mu}_{2}) 
   - 3(\hat{L}\!\cdot\!\vec{\mu}_{1})(\hat{L}\!\cdot\!\vec{\mu}_{2})\Big], \\
\bar{\mathcal{G}}(\vec{\mu}_{1}, \vec{\mu}_{2}) 
&= \frac{9}{8}\Big[(\vec{\mu}_{1}\!\cdot\!\hat{p})(\vec{\mu}_{2}\!\cdot\!\hat{p}) 
   - (\vec{\mu}_{1}\!\cdot\!\hat{q})(\vec{\mu}_{2}\!\cdot\!\hat{q})\Big].
\end{align}
Eccentricity introduces directional asymmetry in the orbital plane, captured by $\mathcal{G}(\vec{\mu}_{1}, \vec{\mu}_{2})$. This contribution vanishes when the magnetic moments have symmetric in-plane projections. The equation of motion can now be expressed as 
\begin{align}\label{eq:equation-of-motion-rewritten-with-f}
 \biggl ( \frac{d^{2} r}{d\phi ^{2}} \dot{\phi ^{2}} \biggr)  &\eta m r^{4}  -   \biggl ( \frac{dr}{d\phi} \biggr)^{2} \dot{\phi^{2}} ( 2 \eta m r^{3} )  \nonumber \\
&  = \eta m r^{5} \dot{\phi^{2}} -  \eta m^{2} r^{2} - 3\,\bar{\mathcal{F}}_{\mathrm{int}}(\vec{\mu}_{1}, \vec{\mu_{2}}, e)    
\end{align}

In the absence of magnetic fields, the orbital solution for an eccentric system is given by
\begin{equation}\label{eq:r-in-eccentric-orbit}
 r = \frac{ a (1- e^2)} {(1 + e \cos \phi)},
\end{equation}
where $a$ is the semi-major axis and $e$ is the eccentricity of the orbit. We adopt a perturbative approach, treating the magnetic field's effect as corrections to the leading-order dynamics. At leading order, the instantaneous orbital frequency is given by $\dot{\phi} = \omega (1 - e^{2})^{-3/2} ( 1+ e \cos \phi)^{2}$~\citep{Peters-Matthews-PhysRev.131.435,BlakeMoore-PhysRevD.93.124061}, where $\omega$ is the orbit-averaged angular frequency. This expression is used in~\cref{eq:equation-of-motion-rewritten-with-f} to obtain
\begin{align}\label{eq:eomrewritten3}
    &(1- e^{2})^{3} \big( \eta m^{2} r^{2} + 3 \, \bar{\mathcal{F}}_{\mathrm{int}}(\vec{\mu}_{1}, \vec{\mu_{2}}, e) \big) + (1 + e cos \phi)^{4} \, \times \nonumber \\
    &\biggl( \eta m \omega ^ {2} r^{4} \frac{d^{2} r}{d \phi ^{2}} - 2 \eta \omega ^{2} m r^{3} \biggl (\frac {dr} {d\phi} \biggr)^{2} - \eta m \omega^{2} r^{5} \biggr) = 0.
\end{align}
To incorporate magnetic corrections, we consider an ansatz of the form
\begin{align}\label{eq:ansatz}
r = \frac{ a (1- e^2)} {(1 + e\, \cos \phi) } (1 + \widetilde{\gamma} \cdot \delta r),
\end{align}
where $\widetilde\gamma$ is a dimensionless parameter. It is defined as 
\begin{align}\label{eq:gamma}
\widetilde \gamma = \frac{\omega^{4/3}}{m^{8/3}\,\eta} \left\{ \bar{\mathcal{F}} (\vec{\mu}_{1},  \vec{\mu}_{2}) + e^2 \, \bar{\mathcal{G}} (\vec{\mu}_{1}, \vec{\mu}) \right\}.
\end{align}
For convenience, we separate this into two components:
\begin{align}
\gamma = \frac{\omega^{4/3}}{m^{8/3}\eta}\,\bar{\mathcal{F}}(\vec{\mu}_{1},\vec{\mu}_{2}), 
\quad \gamma_{B} = \frac{\omega^{4/3}}{m^{8/3}\eta}\,\bar{\mathcal{G}}(\vec{\mu}_{1},\vec{\mu}_{2}),
\end{align}
such that $\widetilde \gamma = \gamma + e^2 \, \gamma_B $. The function $\delta r$ captures the deviation from an unmagnetized eccentric orbit. Substituting this ansatz into~\cref{eq:eomrewritten3}, we obtain the following differential equation for $\delta r (\phi)$:
\begin{align}
    \label{eq:deltar_diffeqn}
    & - 3 (1 + e \cos\phi )^2  + ( - 1 + e^2)^2 \times \nonumber \\ 
    & \bigg( 3 \delta r + 2 e \sin\phi \, \delta r' - (1 + e \cos\phi ) \delta r'' \biggr) =0      
\end{align}
The above equation can be solved to determine $\delta r$, which depends only on eccentricity $e$ and orbital phase $\phi$. We consider a solution of the form $\delta r(\phi) = \delta r_0(\phi) + e \, \delta r_1(\phi) + e^2 \, \delta r_2(\phi)$. Substituting this into~\cref{eq:deltar}, we expand all terms in a power series of $e$ and retain terms up to $\mathcal{O}(e^2)$. We then solve order by order in $e$ to determine $\delta r_0(\phi)$, $\delta r_1(\phi)$, and $\delta r_2(\phi)$. Thus, we get
\begin{equation}\label{eq:deltar}
\delta r = 1 + \frac{3}{2} e \cos\phi + \frac{1}{28} e^2 \biggl( 77 - 3 \cos(2 \phi) \biggr).   
\end{equation}
Incorporating this perturbative correction, $r$ can finally be written as
\begin{align} \label{eq:r_with_deltar}
r &= \frac{ a (1- e^2)} {(1 + e\, \cos \phi) } \biggl\{ 1 + \gamma \biggl[ 1 + \frac{3}{2} e \cos\phi   \nonumber
\\ & \quad{} + \frac{1}{28} e^2 \biggl( 77 - 3 \cos(2\phi)  \biggr) \biggl] + \, e^2 \gamma_{B} \biggl\} 
\end{align}
Note that without magnetic fields, this expression reduces exactly to the unperturbed eccentric orbit given in~\cref{eq:r-in-eccentric-orbit}.

\section{Gravitational and Electromagnetic Energy Loss in Eccentric Binary Inspirals}\label{section:energy-loss-eqs}

The total energy of the system at an instant during the inspiral is given by \citep{ioka2000gravitational}
\begin{equation}\label{eq:energy-instantaneous}
E = \frac{1}{2} \eta m v^2 - \frac{\eta m^2}{r}
- \frac{\mathcal{F}_{\mathrm{int}}(\vec{\mu_{1}}, \vec{\mu_{2}})}{r^3}.
\end{equation}
We use the following standard relation to compute orbital averages~\citep{padmanabhan2010gravitation}
\begin{align}
\langle X \rangle &\equiv \frac{1}{T} \int_0^T dt\, X \\
&= (1 - e^2)^{3/2} \int_0^{2\pi} \frac{d\phi}{2\pi} (1 + e \cos\phi)^{-2} X(\phi),
\end{align}
where $X(\phi)$ is any quantity and $T$ is the orbital time. In our formalism, we neglect orbital precession due to eccentricity and define the orbit with $\phi$ going from 0 to $2\pi$ in one cycle. Applying this averaging to~\cref{eq:energy-instantaneous}, we get
\begin{equation}
E = - \frac{1}{2} \eta m^{5/3} \omega^{2/3}  \left(1 - 2 \gamma - \frac{13 e^2 \gamma}{2} - 4 e^{2} \, \gamma_{B}   \right), 
\end{equation}
where terms containing $\gamma$ represent the corrections due to magnetic interaction. To determine the energy radiated by the system in gravitational waves, we compute the mass quadrupole moment, $Q_{ij}(t) = \mu \left( x^i x^j - \frac{1}{3} \delta_{ij} \, \vec{x} \cdot \vec{x} \right)$, where $\mu$ is the reduced mass. For planar motion, the coordinates are given by $ x(t) = r(t) \cos\phi(t) $ and  $y(t) = r(t) \sin\phi(t)$, with $r(t)$ implicitly given by \cref{eq:r_with_deltar}. The gravitational-wave energy loss is then computed using the standard quadrupole formula
\begin{equation}
\left\langle \frac{dE}{dt} \right\rangle_{\mathrm{GW}} = - \frac{1}{5} \left\langle \dddot{Q}_{ij} \dddot{Q}^{ij} \right\rangle,
\end{equation}
where $<>$ denotes orbital averaging. This leads to the expression
\begin{align}\label{eq:dEdt_gw}
\left\langle \frac{dE}{dt} \right\rangle_{\mathrm{GW}} = & 
-\frac{32}{5} m^{10/3} \eta^2 \omega^{10/3}  \nonumber \\
& \biggl(1 + \frac{157 e^2}{24} + 4 \gamma + \frac{355 e^2 \gamma}{6} +  4 e^2 \gamma_{B} \biggr),
\end{align}
which includes both eccentricity and magnetic interaction contributions. There is also a coupling term between magnetic interaction and eccentricity. In the absence of magnetic effects ($\gamma=\gamma_{B}=0$), this reduces to the standard gravitational-wave energy loss formula for eccentric binaries~\citep{Peters-Matthews-PhysRev.131.435}. For quasicircular orbits, this reduces to the expression obtained by \cite{ioka2000gravitational}. For symmetric orientations of the magnetic moments in the orbital plane (such as $\vec{\mu}_{1,2}\cdot\hat{p}=\vec{\mu}_{1,2}\cdot\hat{q}$ or $\vec{\mu}_{1,2}\parallel\hat{L}$), the parameter $\gamma_{B} =0 $.

In addition to gravitational-wave emission, the system loses energy as a result of the motion of the magnetic dipoles. The effective dipole moment of the system is defined as
\begin{equation}
\vec{\mu}_{\mathrm{eff}} = \frac{1}{m}(m_2 \vec{\mu}_1 - m_1 \vec{\mu}_2).
\end{equation}
The asymmetry of masses and magnetic moments contributes to a non-zero effective dipole. For a binary with symmetric masses and magnetic fields, this net dipole would be zero. The electromagnetic energy radiated due to the motion of this effective dipole is given by
\begin{align}\label{eq:dEdt_em}
\left\langle \frac{dE}{dt} \right\rangle_{\mathrm{EM}} & =
- \frac{1}{15} m^{2/3} \omega^{14/3} \mu_{\mathrm{eff}}^2 
\left[ \mathcal{F}_0(\alpha) + e^2 \, \mathcal{F}_1(\alpha, \beta) \right], \nonumber \\[2ex]
\mathcal{F}_0(\alpha) &= 3 \sin^2\alpha + 4 \cos^2\alpha, \nonumber \\[2ex]
\mathcal{F}_1(\alpha, \beta) & = 45 \sin^2\alpha + 60 \cos^2\alpha - \frac{1}{4} \cos(2\beta)\sin^2\alpha ,
\end{align}
where $\alpha$ is the angle between $\vec{\mu}_{\mathrm{eff}}$ and $\hat{L}$, and $\beta$ is the azimuthal angle of the dipole in the orbital plane. See Appendix~\ref{app:em-radiation} for more details. For the symmetric orientation of the effective dipole in the orbital plane, that is, $\vec{\mu}_{\mathrm{eff}} \cdot \hat{p} = \vec{\mu}_{\mathrm{eff}} \cdot \hat{q}$, $ \mathcal{F}_1(\alpha, \beta)$ simplifies to $ \mathcal{F}_1(\alpha) \approx 45 \sin^2\alpha + 60 \cos^2\alpha$. The total orbit-averaged energy loss, including both gravitational radiation and electromagnetic dipole emission, is expressed as
\begin{equation} 
\left \langle \frac{dE}{dt} \right \rangle = \left\langle \frac{dE}{dt} \right\rangle_{\mathrm{GW}} + \left\langle \frac{dE}{dt} \right\rangle_{\mathrm{EM}}.
\end{equation}

\vspace{0.1cm}
This cumulative loss from gravitational wave emission (including magnetic effects) and electromagnetic radiation drives the binary inspiral. Our expressions closely agree with \cite{henry2024electromagnetic}, obtained using the PN-MPM formalism for aligned dipoles. Our framework extends to arbitrary dipole orientations and considers their influence in the binary inspiral. Next, we examine the phase evolution of the emitted gravitational wave signal. 

The possible GW--EM conversion associated with the Gertsenshtein--Zel'dovich effect is not included in the present analysis. In a neutron-star binary, the magnetic field of the system can induce graviton--photon conversion (and vice versa) \citep{gertsenshtein1962wave,zel1973electromagnetic}. However, this is a propagation effect, acting primarily on the already emitted radiation during its passage through the magnetized environment and therefore does not directly affect the binary dynamics, the energy-loss rate, or the inspiral phase evolution. For completeness, the conversion efficiency in a uniform magnetic field is given by $P_{GW\to EM} \approx \frac{G \bar{B}^{2}L^{2}}{c^{4}}$. Even for extreme neutron-star magnetic fields ($\bar{B}\sim10^{17} \, \mathrm{G}$) and $L \sim 10^6$ cm (near-surface zone), the maximal conversion efficiency is $ \lesssim 8 \times 10^{-4}$. For more typical magnetar fields ($\bar{B}\sim10^{15} \, \mathrm{G}$), it drops to $\sim 8 \times 10^{-8}$. In realistic environments, inhomogeneities and plasma effects are expected to further suppress the conversion by several additional orders of magnitude, particularly at decihertz-kilohertz frequencies \citep{hong2025theoretical, domcke2025gravitational}. The reverse EM $  \to  $ GW process is similarly inefficient. Thus, the associated GW strain attenuation ($ \delta h/h \simeq \frac12 P_{\rm GW\to EM}  $) is correspondingly small and is not expected to appreciably affect the observed gravitational wave signal. Consequently, energy transfer through GW–EM coupling can be reasonably neglected in the present context.

\section{Dephasing}\label{section:dephasing}

The phase evolution of the gravitational wave signal from a coalescing binary system is governed by the following equation \citep{takatsy2025construction}:
\begin{equation}
\frac{d^2\Psi}{d \omega^2} = \frac{2}{\dot{E}}\frac{dE}{d \omega}.
\end{equation}
where $\Psi$ is the Fourier domain phase of the gravitational wave and $\omega$ is the orbital frequency. This relation follows from the stationary phase approximation. The expressions of total energy and total energy loss obtained in the previous section can be used to compute the phase. Orbital eccentricity also evolves with frequency due to both gravitational wave emission and magnetic field effects. For ease of calculation, we assume that the dominant effect of gravitational wave emission alone governs eccentricity evolution. Consequently, from~\cite{phasing-eccentric-3PN-PhysRevD.93.124061}, we use

\begin{equation}\label{eq:eccentricity-evolution}
e(\omega) = e_{0} \left(\frac{\omega_0}{\omega}\right)^{19/18}
\end{equation}
where $e_{0}$ is the orbital eccentricity at frequency $\omega=\omega_0$. Thus, the phase of the gravitational wave signal for a magnetized neutron star binary is given by
\vspace{-0.01cm}
\begin{widetext}
\begin{align}\label{eq:phase-gw-signal1}
\Psi = \frac{3}{128\, m^{5/3} \eta \, \omega^{5/3}} \Bigg\{
&1 - \frac{2355\, e_0^2\, \omega_0^{19/9}}{1462\, \omega^{19/9}} 
- \frac{100\, \bar{\mathcal{F}}(\vec{\mu}_{1}, \vec{\mu}_{2}) \, \omega^{4/3}}{m^{8/3} \eta}
+ \frac{6030\, e_{0}^2\, \bar{\mathcal{F}}(\vec{\mu}_{1}, \vec{\mu}_{2}) \, \omega_0^{19/9}}{341\, m^{8/3} \eta \, \omega^{7/9}} 
+ \frac{600\, e_{0}^2\, \bar{\mathcal{G}}(\vec{\mu}_{1}, \vec{\mu}_{2}) \, \omega_0^{19/9}}{341\, m^{8/3} \eta \, \omega^{7/9}}
\quad \notag \\
&- \frac{5\, \mu_{\mathrm{eff}}^2\, \omega^{4/3} \mathcal{F}_0(\alpha)}{48\, m^{8/3} \eta^2}
+ \frac{e_0^2\, \mu_{\mathrm{eff}}^2\, \omega_0^{19/9}}{m^{8/3} \eta^2\, \omega^{7/9}} 
\left( \frac{785}{10912} \mathcal{F}_0(\alpha) - \frac{15}{2728} \mathcal{F}_1(\alpha, \beta) \right)
\Bigg\}.
\end{align}
\end{widetext}
This expression is in terms of the binary's orbital frequency, which can be related to the gravitational wave dominant-mode frequency via $f = \omega/\pi$. The magnetic terms that enter the phase evolution induce a shift in the gravitational wave phase relative to the non-magnetized case. It is a 2 PN order effect, as evident from the frequency dependence. The phase expression retains full angular dependence. We adopt the azimuthally symmetric limit where $\mathcal{G}=0$ and $\mathcal{F}(\alpha, \beta) = \mathcal{F}(\alpha)$. This choice reduces the number of angular parameters and simplifies the analytic structure. All subsequent discussions are presented in this setting. In this limit, the total dephasing due to magnetic dipole interactions and electromagnetic radiation losses is given by:
\begin{widetext}
\begin{align}\label{eq:dephasing-equation}
\delta \Psi  = \frac{3}{128\, m^{5/3} \eta \, \omega^{5/3}} \Bigg\{ 
& - \frac{100\, \bar{\mathcal{F}}(\vec{\mu}_1, \vec{\mu}_2) \, \omega^{4/3}}{m^{8/3} \eta}
+ \frac{6030\, e_{0}^2\, \bar{\mathcal{F}}(\vec{\mu}_{1}, \vec{\mu}_{2}) \, \omega_0^{19/9}}{341\, m^{8/3} \eta\, \omega^{7/9}}  \notag   \\
&- \frac{5\, \mu_{\mathrm{eff}}^2\, \omega^{4/3} \mathcal{F}_0(\alpha)}{48\, m^{8/3} \eta^2}
+ 
\frac{e_0^2\, \mu_{\mathrm{eff}}^2\, \omega_0^{19/9}}{m^{8/3} \eta^2\, \omega^{7/9}} 
\left( \frac{785}{10912} \mathcal{F}_0(\alpha) - \frac{15}{2728} \mathcal{F}_1(\alpha) \right)
\Bigg\}.
\end{align}
\end{widetext}
The evolution of the gravitational wave phase of a binary neutron star system is influenced by two primary magnetic effects: (i) magnetic interaction, which depends on the product \( \vec{\mu}_1 \cdot \vec{\mu}_2 \) of the star's magnetic moments and their relative orientation, and (ii) electromagnetic emission, governed by the effective magnetic dipole moment and the system's relative velocity. In addition, the phase evolution contains coupling terms that arise from the interplay between orbital eccentricity and magnetic effects. One such term involves the coupling between eccentricity and magnetic interaction, while another captures the coupling between eccentricity and EM emission. Together, these contributions shape the total magnetic imprint on the gravitational wave signal and can significantly affect the phase evolution, especially in systems with high eccentricity and strong magnetic fields.

\begin{figure}[htbp!]
    \centering
\includegraphics[scale=0.46]{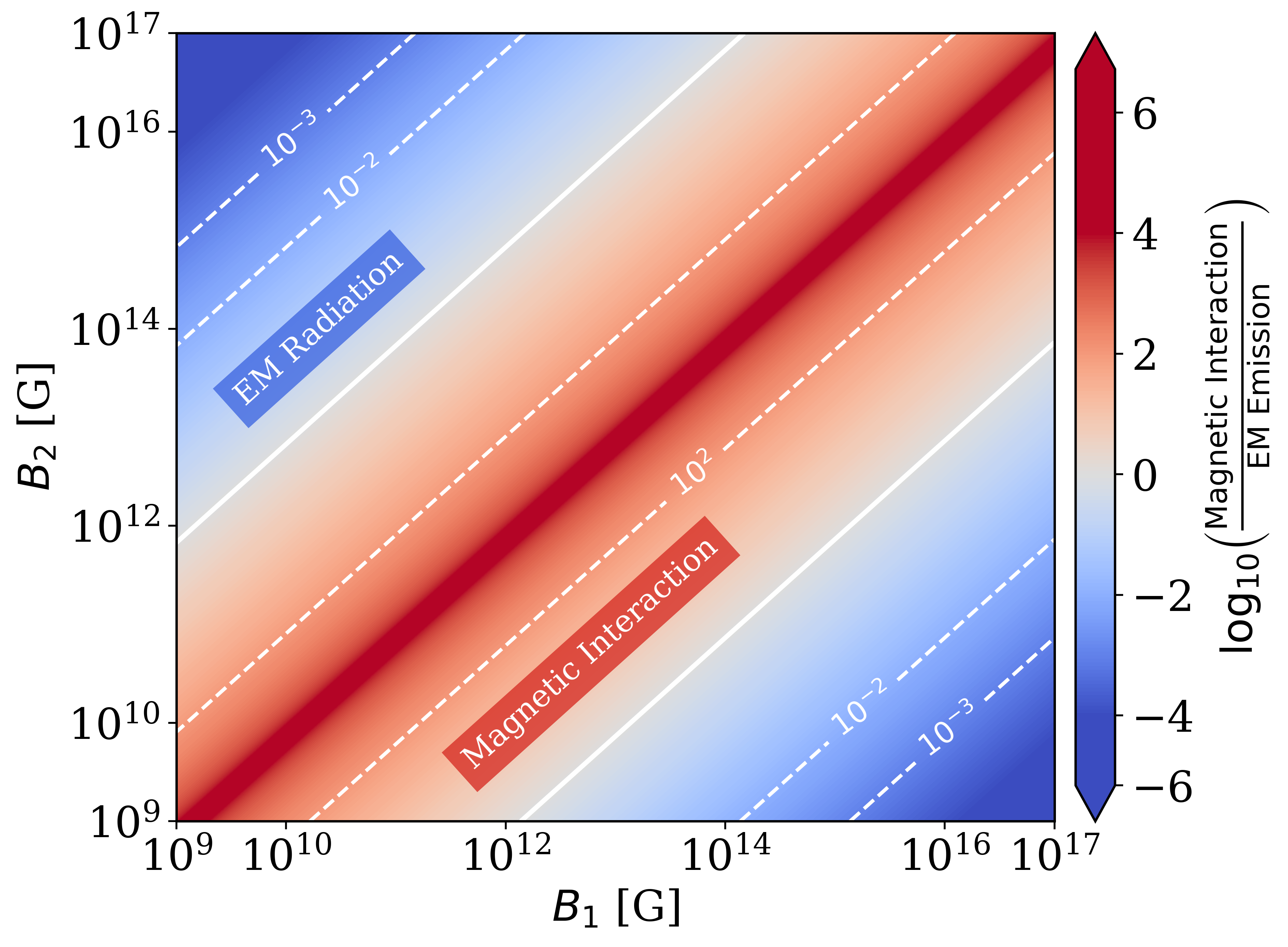}
\caption{The ratio of the magnetic interaction term to the electromagnetic emission term in the gravitational wave dephasing, shown as a function of the surface magnetic field strengths $B_1$ and $B_2$ of the two neutron stars. The case shown is for $m_1 = 2\,M_\odot$ and $m_2 = 1.4\,M_\odot$, assuming a circular orbit ($e = 0$) and $\vec{\mu}_{1,2} \parallel \hat{L}$. The color bar represents $\log_{10}\left( \text{Magnetic Interaction} / \text{EM Emission} \right)$, where blue implies electromagnetic emission dominance and red implies magnetic interaction dominance. The contours (dashed lines) mark the order-of-magnitude levels of this ratio. A transition zone near the white band corresponds to comparable contributions from both terms.}
    \label{fig:dominant_effect}
\end{figure}

\begin{figure*}
    \centering
     \subfloat[\label{fig:corrections_a}]{
     \includegraphics[scale=0.45]{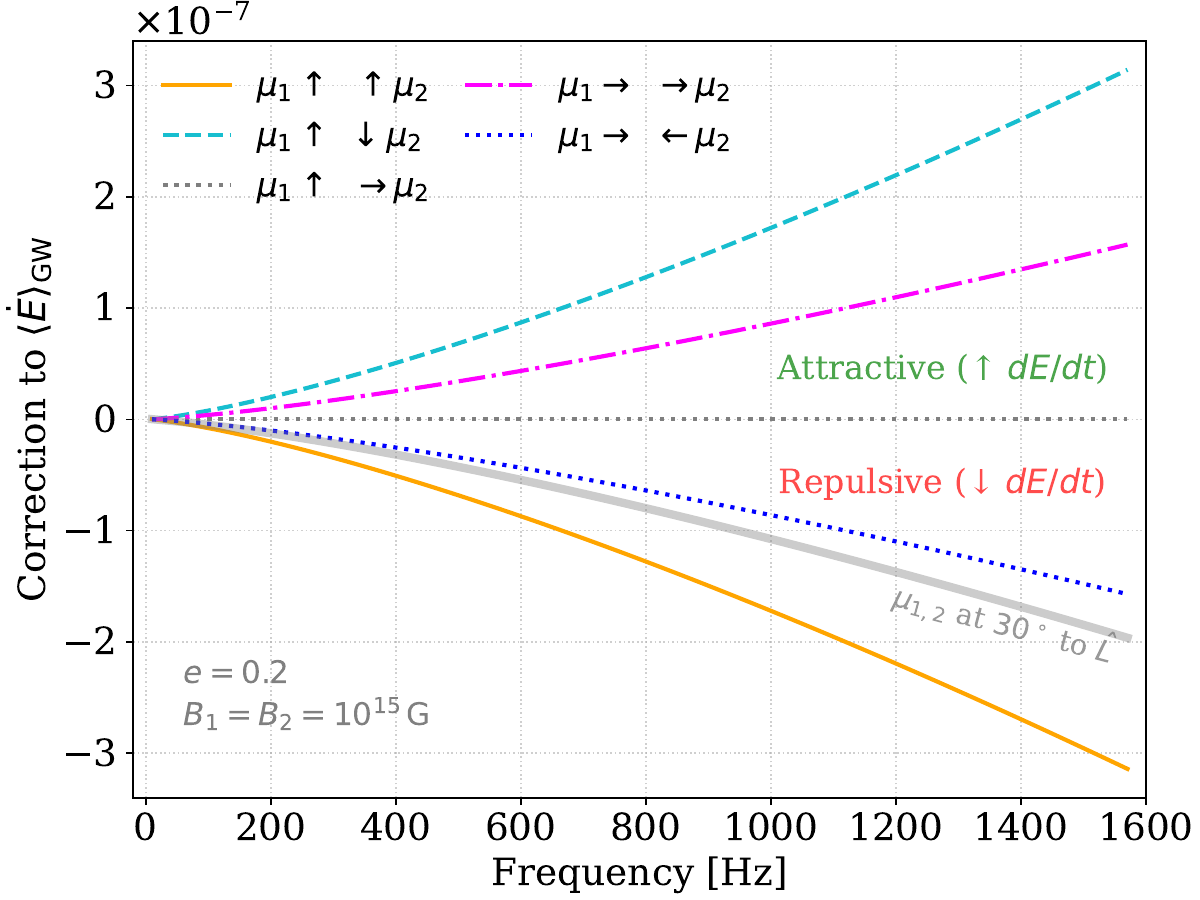}}
    \subfloat[\label{fig:corrections_b}]{\includegraphics[scale=0.45]{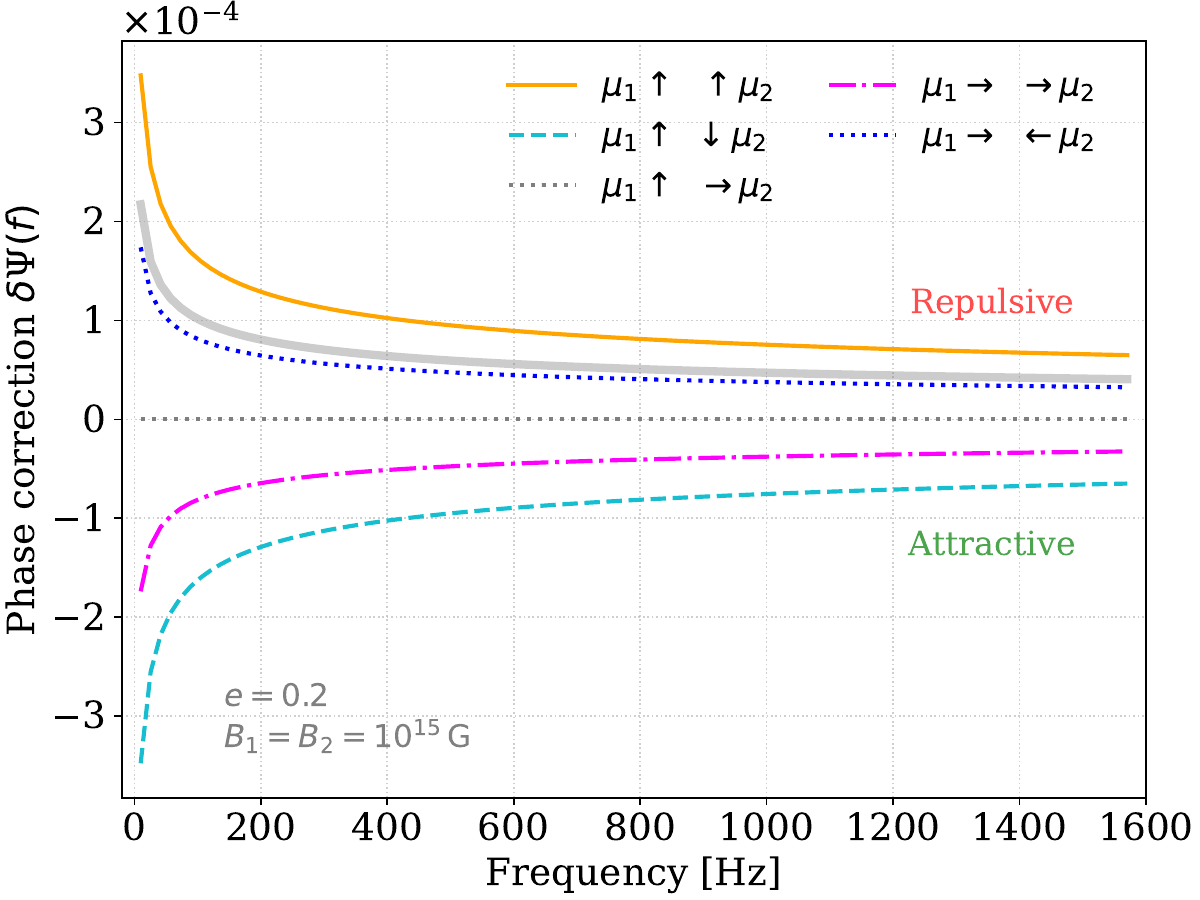}}
    \caption{Magnetic corrections shown as a function of gravitational wave frequency for a binary with component masses $m_1=m_2 = 1.4\, M_{\odot}$, magnetic field strengths $B_{1,2} = 10^{15} \, \mathrm{G}$, and eccentricity $e=0.2$. (a) Fractional correction to gravitational wave energy loss. (b) Phase corrections to the gravitational wave phase. Various orientations of $\vec{\mu}_1$ and $\vec{\mu}_2$ are shown relative to $\hat{L}$ and to each other. Attractive configurations result in faster inspiral and enhance gravitational wave emission, whereas repulsive configurations slow down the inspiral and suppress the gravitational wave emission.}
    \label{fig:magnetic_corrections}
\end{figure*}

\subsection{Dominant magnetic effect}

We explore regions of the BNS parameter space where each effect dominates. This is shown in~\cref{fig:dominant_effect} for a binary with $m_1 = 2\, M_\odot$ and $m_2 = 1.4\,M_\odot$. The plotted quantity is the ratio of the magnetic interaction term to the electromagnetic emission term as it appears in the dephasing expression, assuming a circular orbit ($e = 0$) and orientations $\vec{\mu}_{1, 2}\parallel \hat{L}$. We use the standard relation $\mu = \frac{1}{2} B R^3$ throughout the paper to relate the magnetic moment to the surface magnetic field \citep{ioka2000gravitational}, adopting a fiducial radius of $R = 10 \, \mathrm{km}$ \citep{lattimer2019neutron}. The contours indicate the order of magnitude by which one contribution dominates over the other. Magnetic interaction dominates when both neutron stars possess similar surface magnetic fields, and remains appreciable even if one star's field is lower by one to two orders of magnitude. In contrast, when the magnetic field strengths differ substantially, by several orders of magnitude, the magnetic interaction term becomes negligible, and electromagnetic emission prevails. There also exists a transitional region in the white band where both contributions are of comparable magnitude. Thus, depending on the magnetic field strength and the symmetry of the field configuration within the system, different magnetic mechanisms dominate and govern the imprint on the gravitational wave signal and therefore both must be accounted for consistently.

\subsection{Impact of misaligned orientations}

In the preceding analysis, we assumed that the neutron star magnetic moments were aligned along the orbital angular momentum. However, such alignment might not always occur. Magnetic moments could be aligned in random directions, owing to the capture process and shorter inspiral times. Magnetic interactions are sensitive to the orientation of the dipoles, and such misalignments can subtly influence the inspiral behavior of the system. Figure~\ref{fig:magnetic_corrections} illustrates how different magnetic moment configurations modify the gravitational wave energy loss and phase evolution, for a binary system with $m_1=m_2 = 1.4\, M_{\odot}$, $e=0.2$, and $B_{1,2} = 10^{15} \, \mathrm{G}$. Magnetic corrections to $\langle \dot{E} \rangle_{\mathrm{GW}}$ are presented in \cref{fig:corrections_a}. The plotted quantity is $g(f)= 4\gamma + (355/6) \gamma e^2$, which appears in the expression of energy loss rate. The magnetic interaction potential $ V(r) = - \bar{\mathcal{F}}(\vec{\mu}_{1}, \vec{\mu}_{2})/r^{3}$, depends on the sign of $\bar{\mathcal{F}}(\mu _ {1}, \mu _ {2})$ and determines whether the interaction is attractive or repulsive. An attractive interaction causes a faster inspiral, whereas a repulsive interaction slows down the inspiral. This, in turn, modifies the gravitational wave energy loss at a given frequency. The configurations shown are for magnetic moments $\vec{\mu}_1$ and $\vec{\mu}_2$ parallel or perpendicular to $\hat{L}$, and aligned or anti-aligned with respect to each other. Pulsar observations suggest that magnetic inclination angles range from $5^\circ$ to $87^\circ$, with a population average of $30^{\circ}$ \citep{malov2011angles}. To represent this typical inclination, we include a configuration in which both magnetic moments are inclined at $30^\circ$ to $\hat{L}$. The influence of magnetic alignment is evident. Attractive configurations enhance gravitational wave emission, whereas repulsive configurations suppress it. This magnetic correction to gravitational wave energy loss grows with frequency as the binary inspirals. Intermediate configurations, such as the $30^\circ$ orientations, result in corrections that lie between those from the parallel and perpendicular alignment.

Phase corrections to the gravitational wave signal are shown as a function of frequency for the same orientations (see \cref{fig:corrections_b}). Repulsive configurations result in positive phase shifts at a given frequency, whereas attractive configurations result in negative phase shifts. The second magnetic effect, electromagnetic emission, also contributes to energy loss from the system. It results in gravitational wave phase shifts, as described by \cref{eq:dephasing-equation}. The instantaneous phase shift induced by electromagnetic emission is always negative, since it is governed by the effective dipole $\mu_{\mathrm{eff}}^2$, which remains strictly positive irrespective of the directions of $\vec{\mu}_1$ and $\vec{\mu}_2$. Thus, phase changes from one mechanism may add to or oppose those from the other, depending on orientation. This interplay arises because electromagnetic emission consistently drives the inspiral faster, whereas magnetic interaction may accelerate or decelerate the inspiral depending on the configuration.

\section{Results and Discussion}\label{section:results-discussion}

To assess the contributions of magnetic interaction and electromagnetic emission more precisely, we analyze two fiducial scenarios that isolate each effect.\\

\textit{Case 1: Equal magnetic fields and masses.} Consider a binary system in which both neutron stars possess identical magnetic fields $ B_1 = B_2 = 10^{16} \, \mathrm{G} $ (aligned along $\hat{L}$), and radii $ R \sim 10^6 \, \mathrm{cm}$. This results in magnetic moments 
\begin{equation}
\mu_1 = \mu_2 \sim 10^{33} \, \, \mathrm{G \,cm^3}.    
\end{equation}
Assuming equal component masses ($m_1=m_2=1.4 \, M_\odot$), the effective magnetic dipole moment of the system vanishes, $\mu_{\mathrm{eff}} = 0$. This eliminates the contribution of electromagnetic emission to the system's evolution. However, the magnetic interaction term given by 
\begin{equation}
\bar{\mathcal{F}}(\mu_1, \mu_2) \sim \mu_1 \mu_2 \sim 10^{66} \, \, \mathrm{G^2 \, cm^6}, 
\end{equation}
remains non-zero.\footnote{Here, we refer to the magnitude of $\bar{\mathcal{F}}(\mu_1, \mu_2)$, omitting its sign.} In such symmetric systems, magnetic interaction is the only magnetic effect that influences the orbital evolution and provides the primary magnetic imprint on the gravitational wave signal.\\

\textit{Case 2: Unequal magnetic fields.} Now consider a system in which the neutron stars have significantly different magnetic fields, for instance, $ B_1 = 10^{16} \, \mathrm{G} $ and $ B_2 = 10^{10} \, \mathrm{G}$, such that the magnetic moments are
\begin{equation}
\mu_1 \sim 10^{33} \, \mathrm{G \, cm^3}, \quad \mu_2 \sim 10^{27} \, \mathrm{G \,cm^3}.
\end{equation}
In this system, the magnetic interaction term reduces to
\begin{equation}
\bar{\mathcal{F}}(\mu_1, \mu_2) \sim \mu_1 \mu_2 \sim 10^{60} \, \mathrm{G^2 \, cm^6},
\end{equation}
which is negligible compared to Case 1. Also, the effective magnetic dipole moment is now non-zero. Since $\mu_{1} \gg \mu_{2}$, the effective dipole can be approximated by $ \mu_\mathrm{{eff}} \approx \frac{m_{2}}{m} \mu_{1}$. The squared magnitude governs the dephasing induced by electromagnetic radiation (refer to~\cref{eq:dephasing-equation}, and is given by
\begin{equation}
\mu_{\mathrm{eff}}^2 \sim \mu_1^2 \sim 10^{66} \, \, \mathrm{G^2 \, cm^6},
\end{equation}
which exceeds the magnetic interaction term in this case. As a result, electromagnetic emission emerges as the dominant magnetic contribution to the gravitational wave phase. These two contrasting scenarios illustrate that the relative importance of magnetic interactions and electromagnetic radiation on waveform dephasing strongly depends on the magnetic field magnitudes and their orientations, as previously illustrated in \cref{fig:dominant_effect}. Mass asymmetry ($ m_1\neq m_2$) also enhances electromagnetic emission by inducing an effective dipole moment. Similarly, anti-aligned magnetic moments ($\vec{\mu}_1 = -\vec{\mu}_2$) can also result in a non-vanishing effective dipole. In such configurations, electromagnetic dephasing may act along with magnetic interactions to shape the waveform. However, in this work, we will focus on two idealized regimes where one contribution clearly dominates the other. We now proceed to quantify the dephasing in each case.

\subsection{Dominant effect: Magnetic interaction}

In this subsection, we examine a binary neutron star system with equal masses and magnetic field, i.e., $ m_1 = m_2 = 1.4 \, M_{\odot}$ and $B_1 = B_2$, in an eccentric orbit with eccentricity $e_{0}=0.3$.\footnote{Here, eccentricity of the binary is defined at a gravitational wave frequency of $f=10 \, \mathrm{Hz}$}. The magnetic moments of the neutron stars are assumed to be aligned parallel to the orbital angular momentum. Under these assumptions, the effective magnetic moment of the system vanishes, $ \mu_\mathrm{{eff}} = 0$, and electromagnetic radiation arising from the net dipole in motion would be absent. However, magnetic interactions persist between the neutron stars and influence both the orbital dynamics and the emitted gravitational wave signal.

\begin{figure*}[h!t]
    \centering
    \subfloat[\label{fig:dephasing_a}]{
    \includegraphics[scale=0.45]{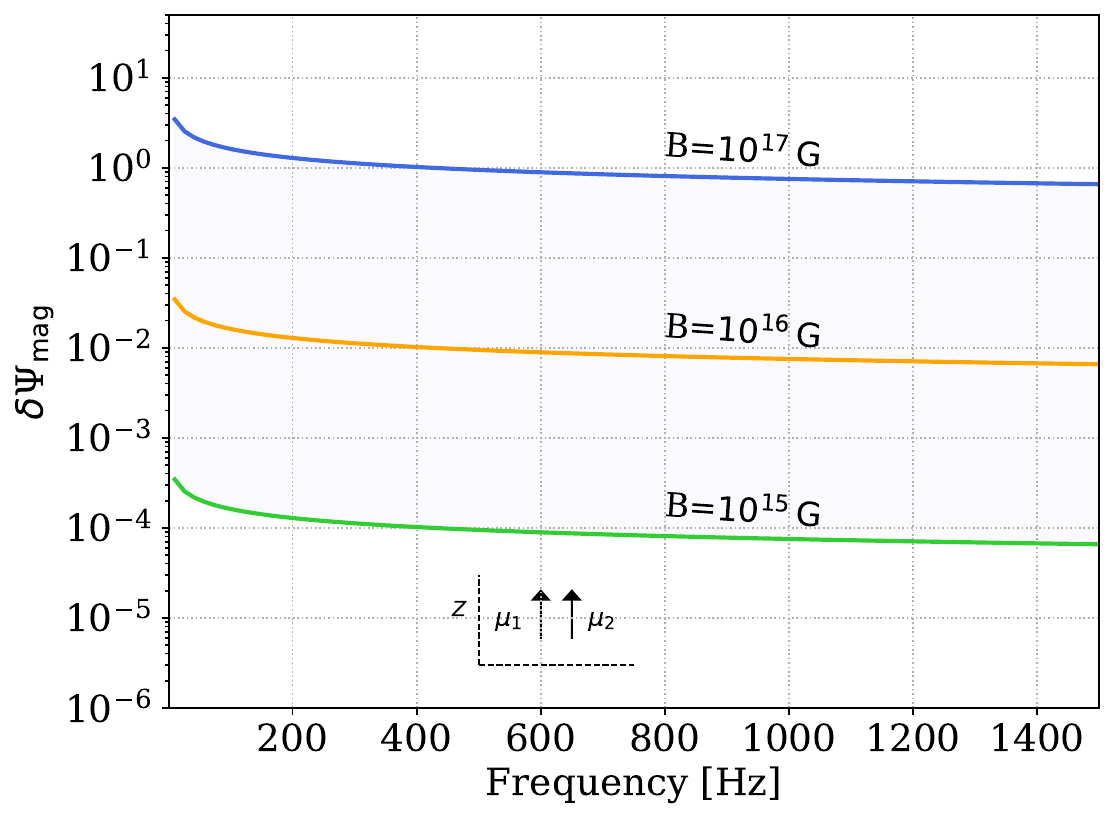}}\quad
    \subfloat[\label{fig:dephasing_b}]{
    \includegraphics[scale=0.45]{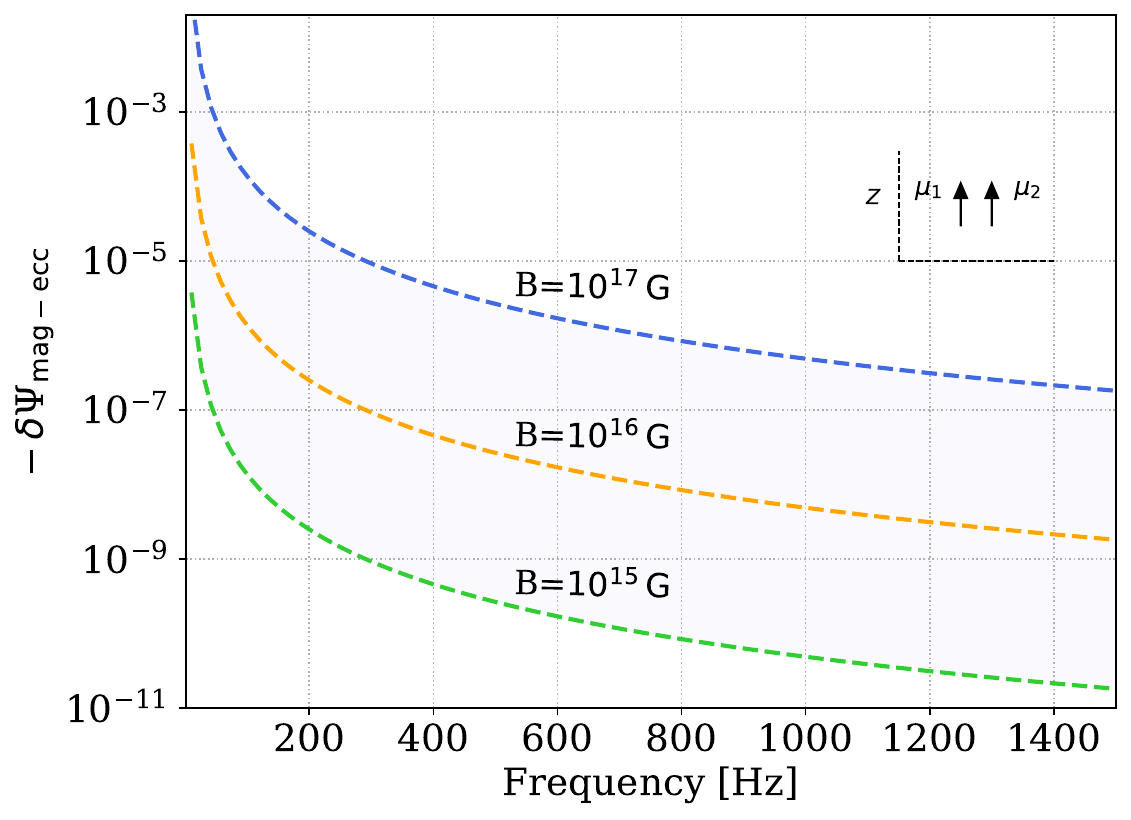}}
   \caption{(a) Total dephasing due to magnetic interaction in an eccentric binary neutron star system with component masses $m_1 = m_2 = 1.4 \, M_{\odot}$ and initial eccentricity $e_0 = 0.3$ (defined at $f = 10 \, \mathrm{Hz}$). The total includes contributions from both the circular magnetic term (non-eccentric) and the eccentricity--magnetic coupling. (b) Dephasing arising solely from the eccentricity--magnetic coupling, shown as a function of surface magnetic field strength ranging from $B = 10^{15}$ to $10^{17}\,\mathrm{G}$. Magnetic moments are assumed to be aligned with the orbital angular momentum. In this configuration, $\vec{\mu}_1 = \vec{\mu}_2$ and hence the effective dipole vanishes $\vec{\mu}_{\mathrm{eff}} = 0.$}
    \label{fig:dephasing}
\end{figure*}

~\cref{fig:dephasing} illustrates gravitational wave dephasing as a function of the gravitational wave frequency for different magnetic fields over a frequency range $f \in [10 \, \mathrm{Hz}, \, f_{\mathrm{ISCO}}]$, covering the sensitivity band of ground-based detectors. We adopt $ f_{\mathrm{ISCO}} = \left( 6^{3/2} \pi (m_1 + m_2) \right)^{-1} $ for the gravitational wave frequency at the innermost stable circular orbit. In phase evolution, the total magnetic interaction comprises two distinct components: the circular magnetic term (non-eccentric) and the eccentricity–magnetic coupling term. These two terms contribute with opposite signs, as described in~\cref{eq:dephasing-equation}. The coupling term cancels out some part of the circular magnetic term to correct for the influence of orbital eccentricity. 

The combined dephasing due to magnetic interactions is shown in~\cref{fig:dephasing_a}, while the isolated contribution of the eccentricity–magnetic coupling term is presented in~\cref{fig:dephasing_b}. As expected, stronger magnetic fields in neutron stars result in greater dephasing in the gravitational wave signal. Specifically, the dephasing is most pronounced at $10^{17}$ G, while it remains relatively small for fields in the range of $10^{15}$ to $10^{16}$ G. For magnetic fields of $10^{14}$ G or lower, the dephasing effect becomes strongly suppressed. Also, the magnetic interaction-induced dephasing decreases with increasing frequency, consistent with its appearance as a 2PN order correction. This behavior differs from tidal interactions, which enter in 5PN order and produce a dephasing that increases with increasing frequency ~\citep{hinderer2010tidal, flanagan2008constraining}. The shaded regions between the curves represent the expected dephasing range for magnetic fields between $10^{15}$ and $10^{17} \, \mathrm{G}$.

The eccentricity-magnetic coupling term, shown in~\cref{fig:dephasing_b}, is several orders of magnitude weaker than the total magnetic interaction contribution (which is dominated by the circular term) and decays rapidly with increasing frequency. This steep decline is a consequence of the decay of orbital eccentricity $e_0$ due to circularization within the observation band. As a result, the coupling term would primarily affect the early inspiral phase and becomes negligible in the final cycles preceding the merger. Although the instantaneous contribution at each frequency is very small, it can accumulate over the duration of the inspiral. 

\begin{figure*}
    \centering
    \subfloat[\label{fig:optimal_snr_magnetic_interaction_a}]{
    \includegraphics[scale=0.45]{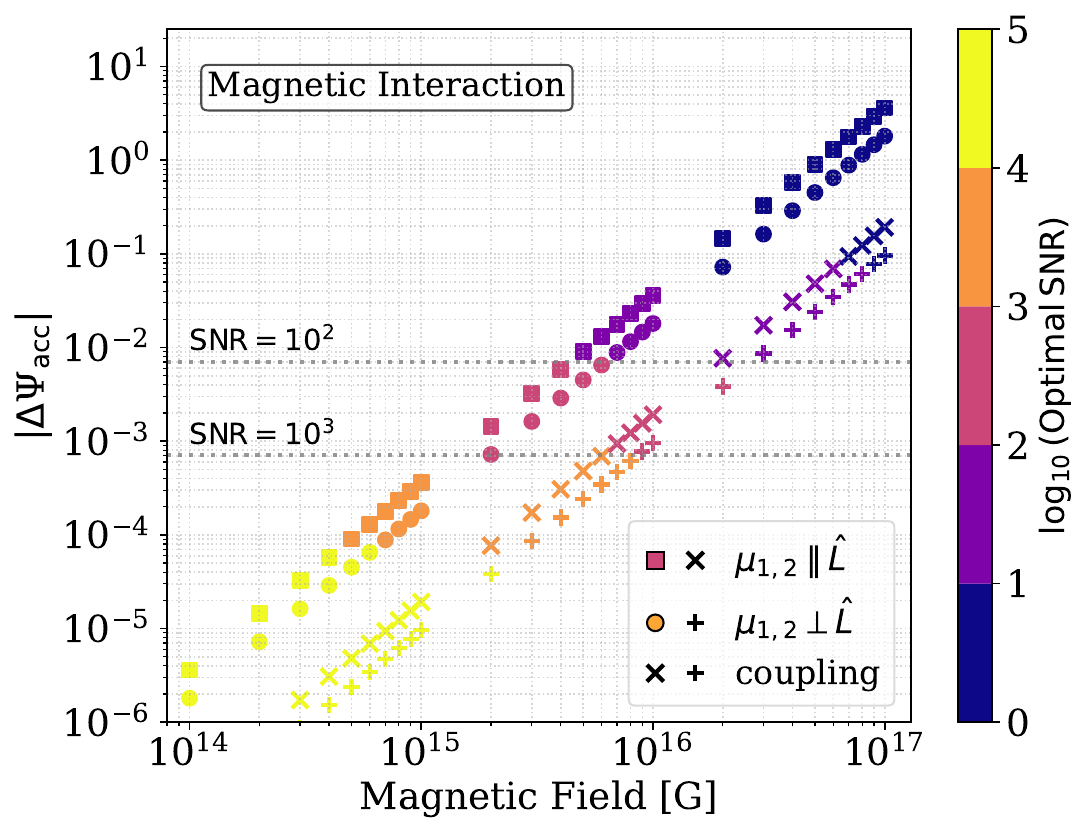}}\quad
    \subfloat[\label{fig:optimal_snr_magnetic_interaction_b}]{
      \includegraphics[scale=0.45]{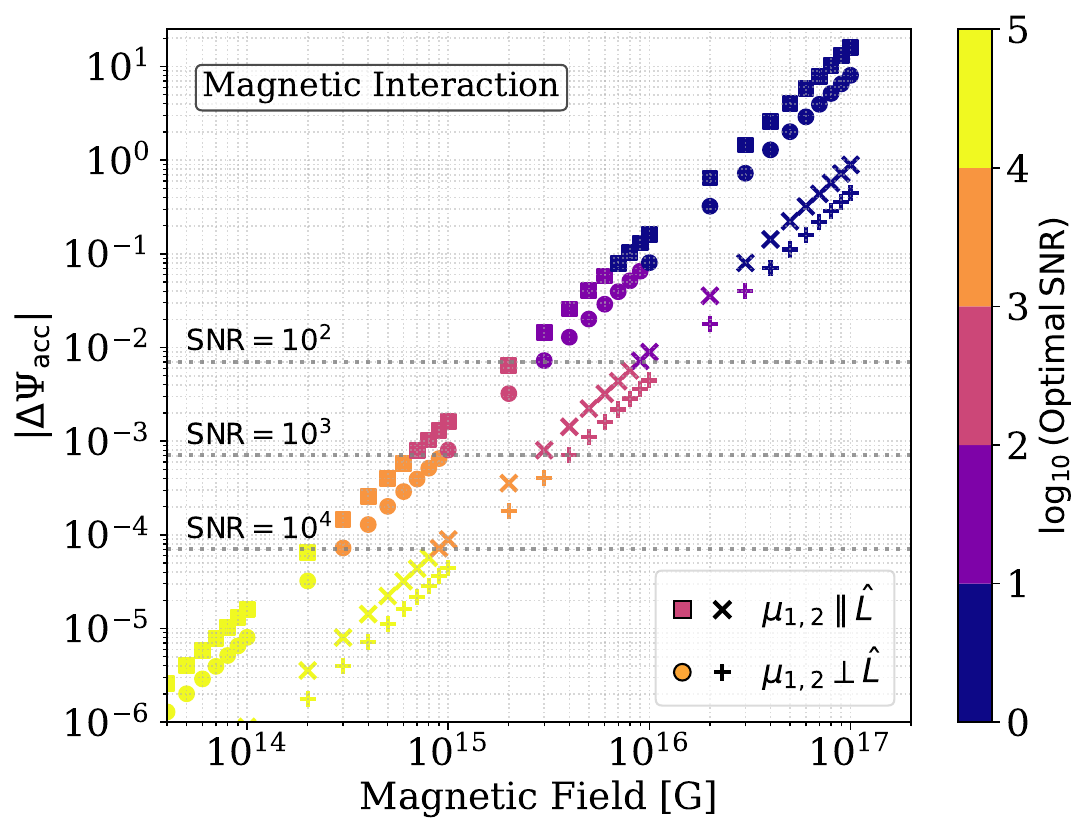}}
   \caption{Optimal signal-to-noise ratio (SNR) required to distinguish the dephasing due to magnetic interaction and eccentricity–magnetic coupling in a binary neutron star system with $m_1 = m_2 = 1.4 \, M_\odot$ and $e_0 = 0.3$, as a function of magnetic field strength. The accumulated dephasing is computed over the frequency range from $f_{\rm start} = 10\,\mathrm{Hz}$ to $f_{\rm ISCO}$ in the left panel (relevant for LIGO, ET, CE), and from $f_{\rm start} = 10^{-1}\,\mathrm{Hz}$ to $10\,\mathrm{Hz}$ in the right panel (DECIGO band). Color bar shows $\log_{10}$(SNR) values required for distinguishability, computed using \cref{eq:optimal-snr}; the dashed lines indicate reference SNR values.}
    \label{fig:optimal_snr_magnetic_interaction}
\end{figure*}

We compute the accumulated dephasing from the point at which the signal enters the detector band at $f_{\text{start}} = 10\,\text{Hz}$ to the end of the inspiral at $f_{\text{end}} = f_{\text{ISCO}}$. To calculate accumulated dephasing, we employ \citep{takatsy2025construction, bernaldez2023eccentricity}
\begin{equation}
\Delta \Psi_{\mathrm{acc}} = \int_{f_{\mathrm{start}}}^{f_{\mathrm{ISCO}}} f \, df \, \frac{d^2 \delta \Psi(f)}{df^2}.
\end{equation}
We focus on two orientations: when the magnetic moments are aligned parallel to $\hat{L}$, and when they are oriented perpendicular to $\hat{L}$. 
The magnetic moment interaction function 
\begin{equation}
\bar{\mathcal{F}}(\vec{\mu}_{1}, \vec{\mu}_{2}) = \frac{1}{2} \left\{ (\vec{\mu}_{1} \cdot \vec{\mu}_{2}) - 3 (\hat{L} \cdot \vec{\mu}_{1})(\hat{L} \cdot \vec{\mu}_{2}) \right\},   
\end{equation}
simplifies to $ \bar{\mathcal{F}}= -(\vec{\mu}_{1} \cdot \vec{\mu}_{2})$ when magnetic moments are aligned parallel to $\hat{L}$, and to $ \bar{\mathcal{F}}= \frac{1}{2}(\vec{\mu}_{1} \cdot \vec{\mu}_{2})$ when magnetic moments are aligned perpendicular. We present the modulus of accumulated dephasing for comparison (refer ~\cref{fig:optimal_snr_magnetic_interaction_a}). The parallel configuration leads to significantly greater dephasing than the perpendicular configuration, as can be seen from the figure. Additionally, we have shown the accumulated dephasing resulting solely from the eccentricity-magnetic coupling. Although this term appears several orders of magnitude smaller in the instantaneous dephasing plot, its contribution to the accumulated dephasing is notable. It is roughly an order of magnitude smaller than the total (circular) magnetic interaction term. Hence, including this coupling term is essential for accurately modeling the gravitational wave phase of eccentric magnetized binaries. 

The optimal signal-to-noise ratio (SNR) required to distinguish the presence or absence of an effect in the measurement can be estimated using~\citep{flanagan1998measuring, lindblom2008model, bernaldez2023eccentricity}
\begin{equation} \label{eq:optimal-snr}
\text{SNR}_{\text{opt}} \approx \frac{1}{\sqrt{ 2 \Delta \Psi_{\mathrm{acc}}^{2}}},    
\end{equation}
where \( \Delta \Psi_{\mathrm{acc}} \) denotes the total accumulated dephasing induced by the effect \footnote{The detectability condition $\overline{\delta\chi}^{\,2}+\overline{\delta\Phi}^{\,2}>1/\rho^{2}$, introduced in Ref.~\citep{lindblom2008model}, requires both the waveform and detector noise for its evaluation. The same work also presents a simpler waveform-based criterion $(\max|\delta\chi|)^2 + (\max|\delta\Phi|)^2 > 1/\rho^2$. In the absence of amplitude corrections, this reduces to $\Delta\Psi_{\mathrm{acc}}^2 > 1/\rho^2$. In our analysis, we employ $2\,\Delta\Psi_{\mathrm{acc}}^2 > 1/\rho^2$ to heuristically account for the amplitude corrections.}. Although this equation is approximate, it provides a reasonably good estimate for detectability. The optimal SNR's are shown in colorbar in~\cref{fig:optimal_snr_magnetic_interaction_a}, with the contours for $\text{SNR} = 100$ and $\text{SNR} = 1000$ indicated.

The parallel configuration is associated with stronger accumulated dephasing and requires a lower SNR threshold for detection compared to the perpendicular alignment. Thus, for a fixed SNR, weaker magnetic fields can be distinguished in the parallel case, as evident in the figure. An optimal SNR of approximately 100 is required to distinguish magnetic fields of $10^{16}\,\mathrm{G}$ in the gravitational wave signal, whereas detecting fields of $10^{15}\,\mathrm{G}$ requires an SNR of 1000. Such high SNRs are expected to be routinely achievable in the third-generation era with detectors such as ET and CE. For example, an optimal event similar to GW170817 observed with ET is predicted to have an SNR of $\sim1750$~\citep{jimenez2018impact}. Hence, next-generation detectors could uncover the subtle influence of strong magnetic fields in the gravitational waves emitted by neutron star binaries. The parallel and perpendicular configurations shown are two extreme limits, and the accumulated dephasing and optimal SNR for intermediate orientations are expected to be between them. Although the coupling terms are more subtle, they could still be resolved at very high SNRs, provided that the magnetic fields inducing them are sufficiently strong.

\begin{figure*}
    \centering
    \subfloat[\label{fig:optimal_snr_em_emission_a}]{
    \includegraphics[scale=0.45]{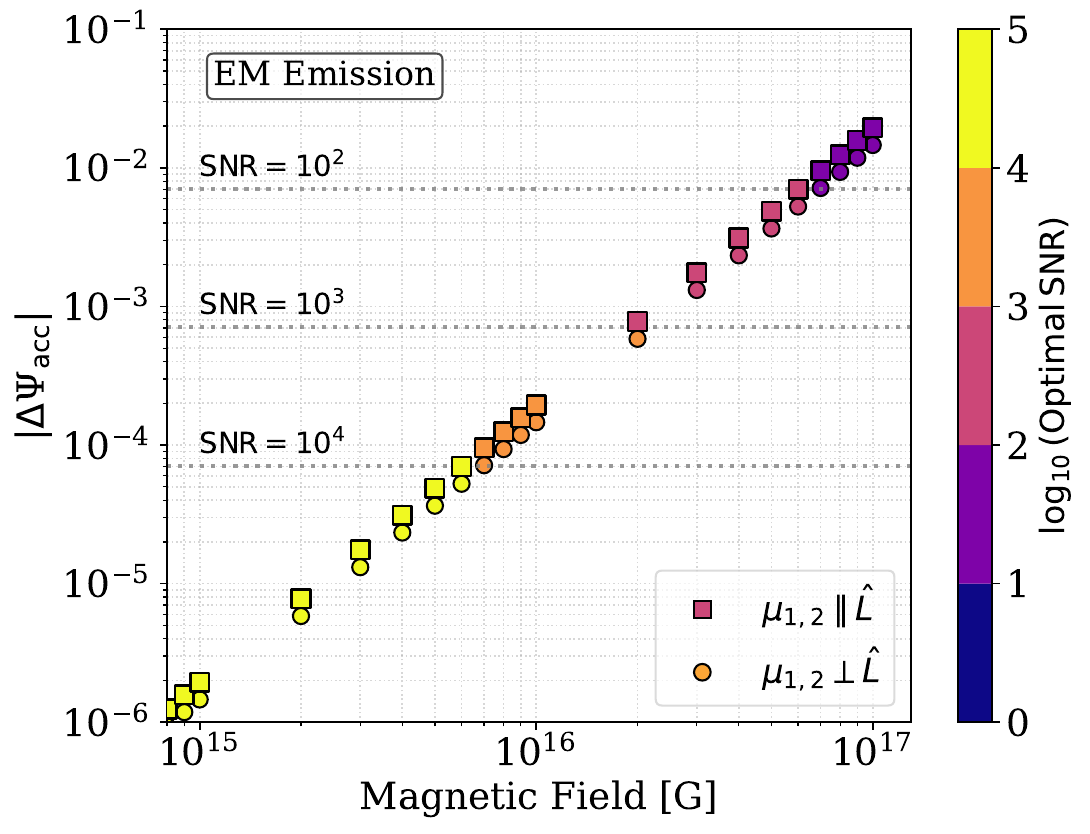}} \quad
    \subfloat[\label{fig:optimal_snr_em_emission_b}]{
      \includegraphics[scale=0.45]{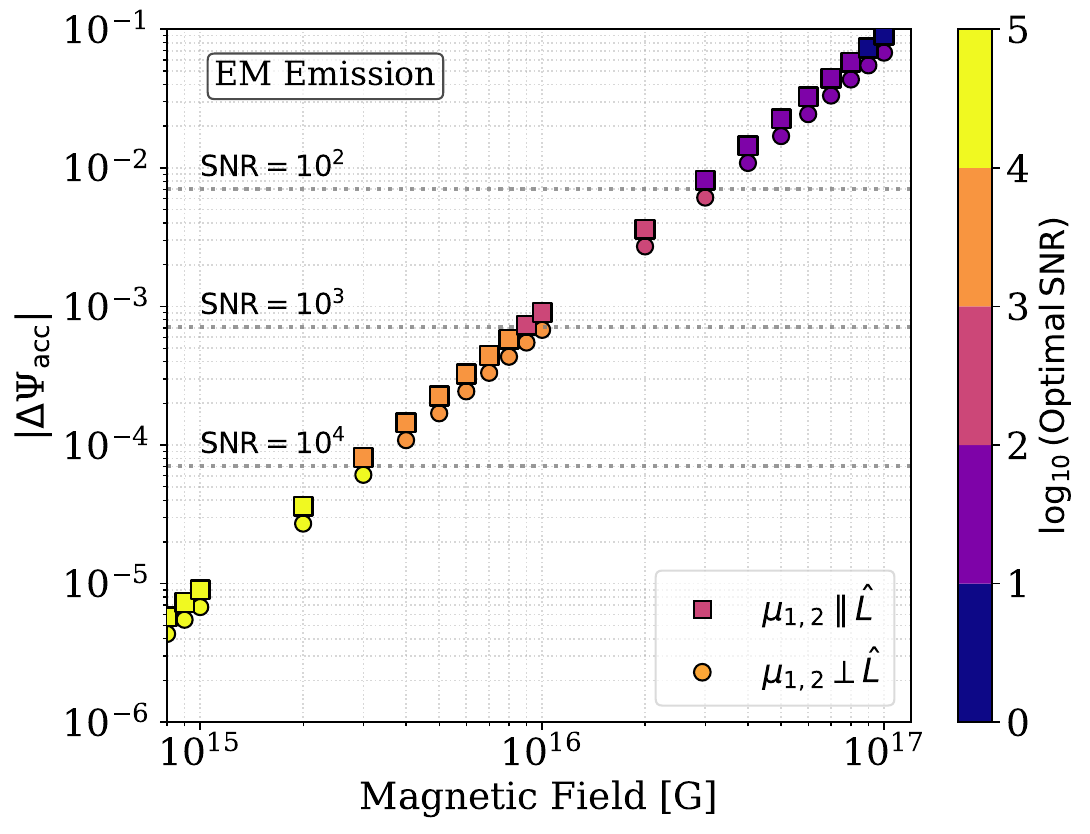}}
    \caption{Optimal signal-to-noise ratio (SNR) required to detect dephasing due to electromagnetic dipole emission in an eccentric binary neutron star system with $m_1 = m_2 = 1.4 \, M_\odot$, $e_0 = 0.3$, and $B_2 = 10^{10}$ G, as a function of $B_1$. The accumulated dephasing is computed from $f_{\rm start} = 10\,\mathrm{Hz}$ to $f_{\rm ISCO}$ in the left panel (LIGO, ET, CE band), and from $f_{\rm start} = 10^{-1}\,\mathrm{Hz}$ to $10\,\mathrm{Hz}$ in the right panel (DECIGO band). 
Color bar indicates $\log_{10}(\mathrm{SNR})$ values required for distinguishability, calculated using \cref{eq:optimal-snr}. Dashed lines correspond to the reference SNR levels of 100, 1000, and 10000.
}
    \label{fig:optimal_snr_em_emission}
\end{figure*}

In~\cref{fig:optimal_snr_magnetic_interaction_b}, we present the accumulated dephasing and corresponding optimal SNR, for the sensitivity band of DECIGO, with bandwidth $[f_{\mathrm{start}}, \, f_{\mathrm{end}}] = [10^{-1}\, \mathrm{Hz}, \, 10 \, \mathrm{Hz}]$. The eccentricity is set to $e_{0}=0.3$ at $f_{\mathrm{start}} = 10^{-1} \, \mathrm{Hz}$. DECIGO is a proposed space-based gravitational wave observatory that will operate in the decihertz band. Its frequency band spans nearly three orders of magnitude, and gravitational waves from coalescing binaries can be tracked for hundreds of days or several months, resulting in significantly high SNRs. With an SNR of 1000, magnetic fields of a few times $10^{14}\,\mathrm{G}$ can be distinguished. While an SNR of 10,000 would enable sensitivity to magnetic fields as low as $10^{14}\,\mathrm{G}$. Thus, gravitational waves offer a promising avenue to probe the population of magnetar binaries. Stronger magnetic fields lead to more pronounced dephasing signatures, making them easier to detect. The trends observed for magnetic dipole orientation and eccentricity-magnetic coupling resemble those seen in the ground-based case: the parallel configuration results in stronger dephasing and lower SNR thresholds, whereas coupling-induced effects remain fainter and resolvable at sufficiently high SNRs. The presence of ultra-strong fields $ \sim 10^{16–17} \mathrm{G}$ would produce a particularly pronounced phase signature, offering a valuable observational opportunity.

When both neutron stars are equally and strongly magnetized, for example, each with a magnetic field of $10^{16}\,\mathrm{G}$, the magnetic interactions are significantly enhanced. However, reducing the magnetic field strength of one star by even an order of magnitude, such as from $10^{16}\,\mathrm{G}$ to $10^{15}\,\mathrm{G}$, leads to a noticeable weakening of the interaction, roughly by a similar factor. Although gravitational wave dephasing and SNR$_\mathrm{opt}$ for such mildly asymmetric configurations are not explored here, they likely represent more probable and astrophysically realistic systems. They occupy an intermediate regime between the equal-field binaries analyzed here and the more highly asymmetric cases in the next section.

\subsection{Dominant effect: Electromagnetic emission}
In the previous section, we examined binaries for which the effective dipole moment vanishes and only magnetic interactions shape the gravitational wave signal. We now turn to systems with a pronounced magnetic field asymmetry. In these cases, the primary neutron star's magnetic field is much stronger than that of the companion, creating conditions for electromagnetic emission and introducing an additional channel for gravitational wave phase modulation.

We consider a binary with component masses and magnetic fields given by $m_1 = m_2 = 1.4 \, M_{\odot}$, $B_1 \gg B_2$, and $B_2 = 10^{10} \, \mathrm{G}$. The accumulated dephasing and corresponding optimal SNR are shown in~\cref{fig:optimal_snr_em_emission_a} as functions of the primary neutron star's magnetic field $B_1$. For sources in the frequency band of ground-based observatories (such as LIGO, ET, and CE), the dephasing is integrated over the frequency range $f \in [10\, \mathrm{Hz},\, f_{\mathrm{ISCO}}]$. For DECIGO-band sources, the integration spans $f \in [10^{-1}\, \mathrm{Hz},\, 10\, \mathrm{Hz}]$, as shown in~\cref{fig:optimal_snr_em_emission_b}. The initial eccentricity $e_0 =0.3$ is defined at the lower end of each respective band. The configurations illustrated involve $\vec{\mu}_{1,2} \parallel \hat{L}$ and $\vec{\mu}_{1,2} \perp \hat{L}$, which represent two distinctive alignments relative to the orbital angular momentum. In this regime dominated by the effective dipole of the system, magnetic orientations do not significantly influence the SNR required for detectability. The parallel and perpendicular configurations produce nearly identical optimal SNR curves. Our results indicate that, for an SNR of 100, magnetic fields become distinguishable only if $B_1 \gtrsim 10^{17} \, \mathrm{G}$. For $\mathrm{SNR} = 1000$, ground-based detectors can detect fields of $B_1 \approx 2 \times 10^{16} \, \mathrm{G}$. In the DECIGO band, fields as low as $B_1 \approx 10^{15} \, \mathrm{G}$ could be resolved in ultra-high SNR events ($\mathrm{SNR} \sim 10^4$), while fields of $ \sim 10^{16} \, \mathrm{G}$ become distinguishable for SNRs of 1000, as well as 100. Thus, binaries consisting of a magnetar and a weakly magnetized companion may leave measurable imprints on gravitational wave signals via effective dipole-driven emission. Under favorable signal-to-noise conditions, next-generation detectors could probe such systems.

Electromagnetic dephasing can also arise in NSBH systems. In these binaries, the black hole is essentially non-magnetized and its magnetic moment is zero ($\vec{\mu}_1 = 0$), while the neutron star has a finite magnetic moment ($\vec{\mu}_2$). This difference can create an effective dipole that can influence the orbital dynamics via the emission of electromagnetic radiation. The gravitational waves emitted during inspiral from such systems may carry phase shifts linked to this asymmetry. The NSBH inspirals typically end at lower gravitational wave frequencies than BNS systems, which could make the dephasing acquired over the signal duration smaller. However, the high SNRs associated with NSBH systems may compensate for the reduced dephasing and enhance detectability. The present formalism can describe such systems, serving as a probe of the companion neutron star's magnetic field.

\subsection{Distinguishability and Measurement Prospects}
So far, we have estimated the accumulated dephasing and examined its detectability using the optimal SNR. While this provides a reasonable estimate, we now shift to a more comprehensive analysis that incorporates both the waveform and the detector response. We consider two gravitational waveforms: let $h(f)$ denote the waveform without magnetic field effects, and $h_{\mathrm{mag}}(f)$ denote the waveform that includes magnetic effects. The difference between the two is 
\begin{equation}
\delta h = h_{\mathrm{mag}} - h .
\end{equation}
The quantity $\delta h$ captures both the phase shifts and the amplitude changes that arise from magnetic effects. A gravitational wave detector can distinguish the two signals if the norm of this difference satisfies the criterion~\citep{lindblom2008model}  
\begin{equation}
\langle \delta h \mid \delta h \rangle > 1 .
\end{equation}
The above condition ensures that the waveform deviations caused by the magnetic fields are measurable above the noise floor of a given gravitational wave detector. The inner product between two waveforms $a$ and $b$ is defined as
\begin{equation}    
\langle a | b \rangle = 4 \, \mathrm{Re} \int_0^\infty \frac{\tilde{a}(f) \tilde{b}^*(f)}{S_n(f)} \, \mathrm{d}f,
\end{equation}
where $\tilde{a}(f)$ and $\tilde{b}(f)$ are the waveforms in the frequency domain and $S_n(f)$ is the one-sided power spectral density of the detector noise. Since our systems of interest are eccentric binaries, we employ a frequency-domain waveform model that incorporates orbital eccentricity as described in Refs.~\cite{yunes2009post, huerta2014accurate}. This model accounts for the subdominant harmonics and provides an accurate representation of the gravitational wave signal over the relevant frequency range. We take this waveform as our reference signal $h(f)$. By incorporating the magnetic dephasing expressions obtained in the previous section, we construct a modified waveform $h_{\mathrm{mag}}(f)$ that characterizes magnetized eccentric binaries. For the detector response, we use the O5 sensitivity curve for LIGO, and the design sensitivity curves for ET and DECIGO~\citep{dupletsa2023gwfish, schmitz2021new}.

\begin{figure}
    \centering
    \includegraphics[scale=0.44]{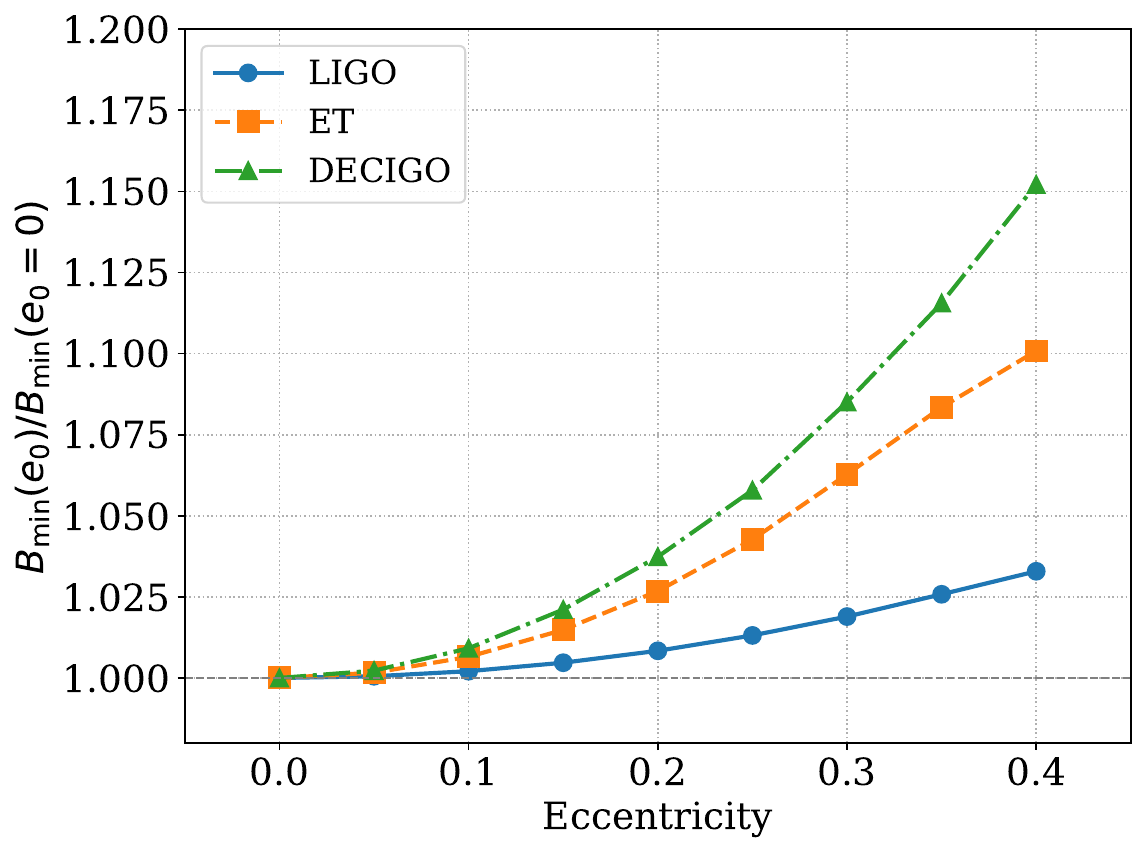}
    \caption{Ratio of the minimum magnetic field $B_{\mathrm{min}}(e_{0})$ required for $\langle \delta h \mid \delta h \rangle = 1$ to its circular-orbit value $B_{\mathrm{min}}(0)$, shown as a function of orbital eccentricity. This ratio increases to $\sim 3\%$ at $e_{0} = 0.4$ for LIGO sources. Hence, eccentric binaries require somewhat stronger magnetic fields than circular binaries for effects to become detectable. For ET and DECIGO, the ratio rises to about $10\%$ and $15\%$ at the same eccentricity.} \label{fig:ratio_magnetic_field_vs_eccentricity}
\end{figure}

\begin{figure*}
\centering
    \subfloat[\label{fig:horizon_distance_a}]{
    \includegraphics[scale=0.44]{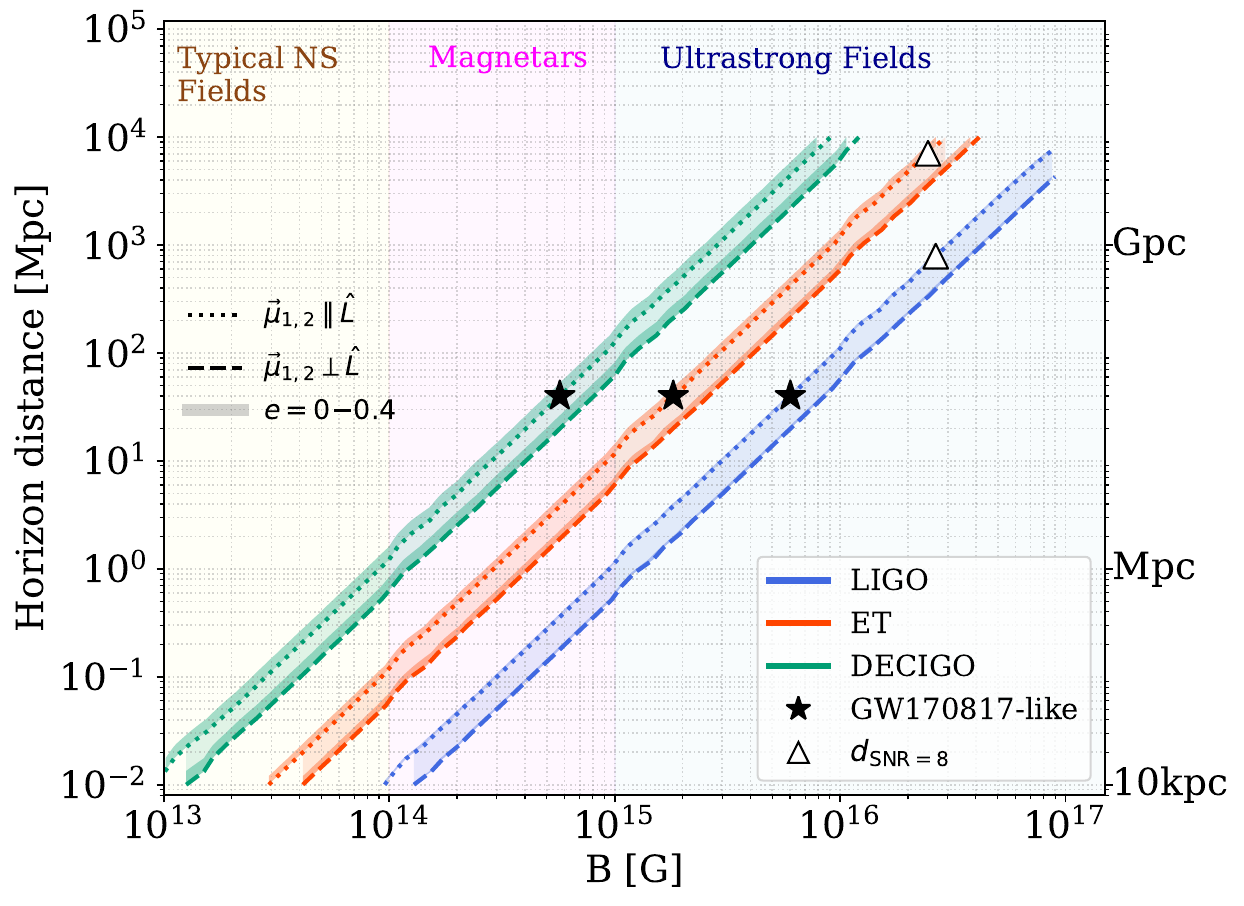}}
    \subfloat[\label{fig:horizon_distance_b}]{
    \includegraphics[scale=0.44]{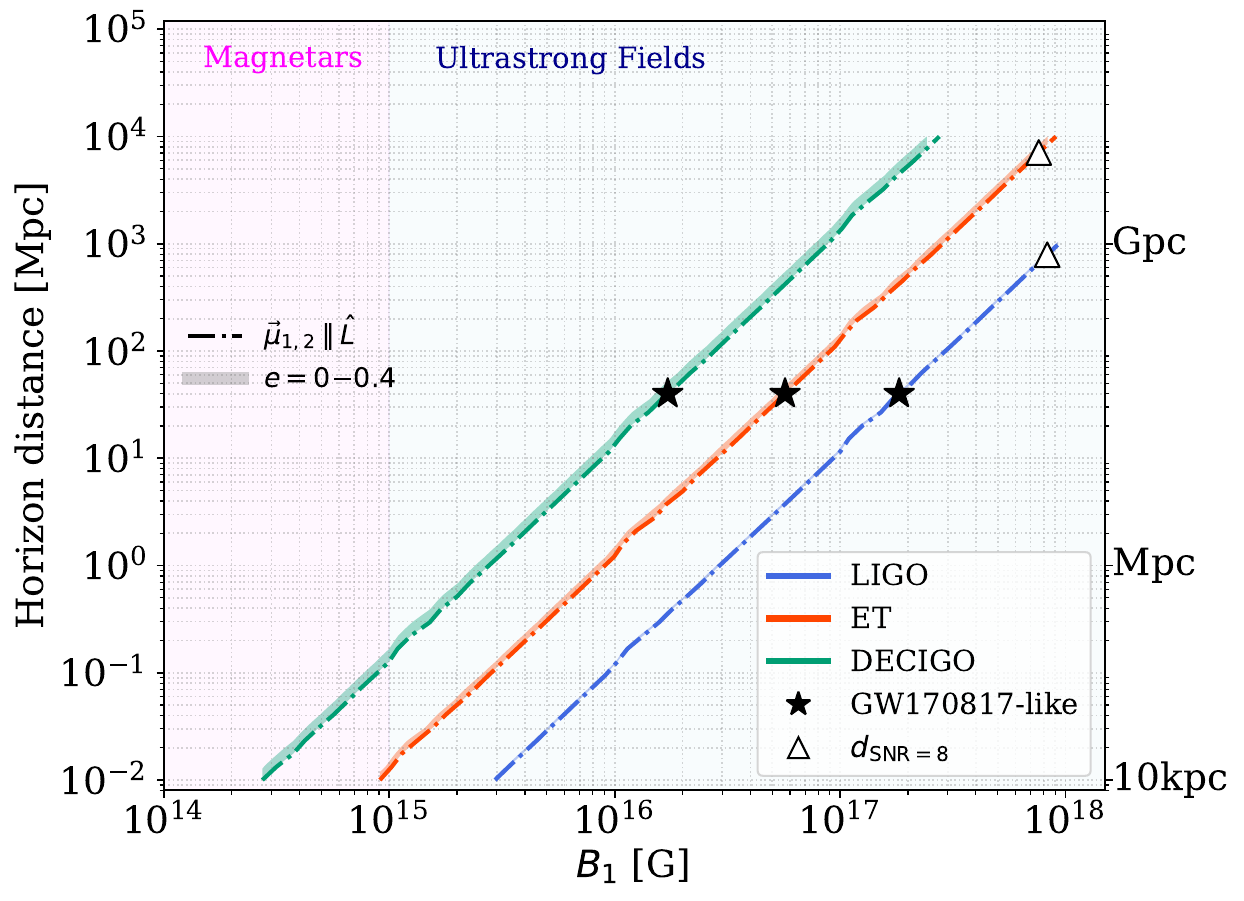}}
     \caption{The horizon distance up to which detectors are sensitive to the presence of magnetic effects in the gravitational waveform is shown as a function of the magnetic field strength for a binary neutron star system with component masses $m_1 = m_2 = 1.4 \, M_\odot$. Initial eccentricity $e_{0}$ is defined at the respective band entry frequency of each detector.
    Two orientations of magnetic moments with respect to orbital angular momentum are shown. For each orientation, the shaded band corresponds to binary eccentricities ranging from 0 to 0.4. Results are shown for LIGO (O5), ET, and DECIGO. (a) shows symmetric case $B_1 = B_2$, and (b) depicts asymmetric case $B_1 \gg B_2 = 10^{10} \, \mathrm{G}$. The  background colors indicate different magnetic field regimes: $<10^{14} \, \mathrm{G}$ (typical neutron stars), $10^{14}$-$10^{15} \, \mathrm{G}$ (magnetars), and $>10^{15} \, \mathrm{G}$ (ultrastrong magnetars).} \label{fig:horizon_distance}
\end{figure*}

We examine a binary neutron star system with equal masses $m_{1} = m_{2} = 1.4 \, M_{\odot}$ located at $d=100 \, \mathrm{Mpc}$. Additionally, we assume equal fields for the neutron stars $B_{1}=B_{2}=B$, aligned along the orbital angular momentum. We vary the magnetic field from $10^{13}$ to $10^{18} \,\mathrm{G}$, and orbital eccentricities in the range $e_{0} \in [0, 0.4]$. We define $e_0$ at the lower end of the detector band, for LIGO at $10$ Hz, ET at $5$ Hz, and DECIGO at $10^{-1}$ Hz. For each pair of magnetic field and eccentricity value $(B, e_{0})$, we compute $ \langle \delta h | \delta h \rangle $. We use the threshold $ \langle \delta h | \delta h \rangle = 1$ to identify the minimum magnetic field strength $ {B_\mathrm{min} (e_{0})}$ required for the magnetic effects to produce a measurable impact on the gravitational wave signal. 

The ratio $B_\mathrm{min}(e_{0})/B_\mathrm{min}(e_{0}=0)$ is illustrated in~\cref{fig:ratio_magnetic_field_vs_eccentricity}. For LIGO sensitivities, eccentric binaries require magnetic field strengths approximately $1-3\%$ higher than those in the circular case to satisfy the distinguishability threshold. This additional field strength is necessary because the eccentricity-magnetic coupling term carries opposite sign compared to the pure magnetic term, a feature that holds across all magnetic orientations. Therefore, a higher magnetic field is needed at larger eccentricities to produce the same level of dephasing for distinguishability. In the band of the ET, binaries enter at comparatively lower frequencies and maintain their eccentricity over extended inspiral durations, resulting in a stronger accumulation of magnetic dephasing. The ratio $ B_\mathrm{min}(e_{0})/B_\mathrm{min}(e_{0}=0) $ reaches $ \sim 10\%$ at $e_{0} \approx 0.4$. A similar trend is observed for DECIGO, with ratio approaching $\sim 15\%$ at $e_{0} \approx 0.4$ \footnote{The ratio depends on the reference frequency $f_{\rm start}$ at which $e_0$ is defined; for example, defining $e_0$ at $f_{\rm start} = 10^{-1} \, \mathrm{Hz}$ versus $10^{-2} \, \mathrm{Hz}$ in DECIGO results in different values. Each $ (e_0, f_{\rm start})$ pair for fixed $(m_1, m_2)$ specifies a physically distinct binary system.}. Thus, highly eccentric systems observed by these detectors must host stronger magnetic fields to leave a measurable imprint on the gravitational wave signal.

We present the horizon distance up to which the presence of magnetic interaction effects in the waveform can be distinguished in~\cref{fig:horizon_distance_a} for three detector configurations: LIGO, ET, and DECIGO. The binary consists of masses $m_1 = m_2 = 1.4 \, M_{\odot}$ and magnetic field strengths $B_{1} = B_{2} = B$ \footnote{We assume optimal orientation and location (overhead) for the binary source.}. The source distance (horizon distance) spans from $10 \, \mathrm{kpc}$ to $10 \, \mathrm{Gpc}$, where $10$ kpc corresponds to a galactic source and $10 \, \mathrm{Gpc}$ corresponds to a redshift of $\mathrm{z \approx 1.6}$. The distances and scales are shown on the vertical axis, and the magnetic field strength is shown on the horizontal axis. Two magnetic orientations are considered: the dotted line represents $\vec{\mu}_{1,2}$ parallel to $\hat{L}$ and the dashed line represents $ \vec{\mu}_{1,2}$ perpendicular to $\hat{L}$. In both cases, $\vec{\mu}_1$ and $\vec{\mu}_{2}$ are mutually aligned. Intermediate orientations are expected to span the region between these two curves, and their effects are effectively captured within this envelope. For each configuration, the shaded region corresponds to eccentricities ranging from $0$ to $0.4$.

For a fixed detector, the parallel configuration yields a greater horizon distance than the perpendicular configuration. Eccentricity introduces small variations in horizon distance, reflected in the narrow bands shown for each configuration. Eccentricity has a negligible impact on the horizon distance for LIGO-band binaries, but it affects the horizon distances of ET and DECIGO-band binaries to a modest extent, as evident from the width of the corresponding bands.  The parallel orientation represents the most optimistic scenario. We find that ET and DECIGO exhibit significantly greater sensitivity to magnetic field effects than LIGO. For galactic binaries, ET and DECIGO can probe magnetic fields as weak as a few times $ 10^{13} \, \mathrm{G}$, while LIGO is limited to fields stronger than $ \sim 10^{14} \, \mathrm{G}$. In terms of spatial reach, both ET and DECIGO have greater reach. They can detect the same magnetic field at significantly greater distances than LIGO. For instance, a magnetic field of $ B\sim 10^{16} \, \mathrm{G}$ is detectable up to $1 \, \mathrm{Gpc}$ in ET and up to $10 \, \mathrm{Gpc}$ with DECIGO, whereas the reach of LIGO is limited to a few $100 \, \mathrm{Mpc}$. 

~\cref{fig:horizon_distance_b} shows the maximum distance at which electromagnetic energy loss in a neutron star binary can be detected in the waveform for three detector configurations. Unlike magnetic interaction, this mechanism becomes relevant when the magnetic-field asymmetry between neutron stars is significant. The source parameters are $m_1 = m_2 = 1.4 \, M_{\odot}$, $B_{2} = 10^{10} \, \mathrm{G}$, and $B_1 \gg B_2$. Since the magnetic field of the primary is much stronger, the direction of $\vec{\mu}_2$ does not affect the electromagnetic emission and thus the horizon distance. We therefore show the results for a single orientation. The source distance again ranges from $10 \, \mathrm{kpc}$ to $10 \, \mathrm{Gpc}$. DECIGO is significantly more sensitive than ET and LIGO, capable of detecting fields as low as $10^{14} \, \mathrm{G}$ for nearby galactic sources. ET is sensitive to fields stronger than $ \sim 10^{15} \mathrm{G}$, whereas LIGO is limited to fields above $3 \times 10^{15} \, \mathrm{G}$. For ultra-strong fields ($ B \sim 3 \times 10^{17} \, \mathrm{G}$), ET and DECIGO can reach as far as $\sim 1-10 \, \mathrm{Gpc}$, whereas LIGO's reach is confined to $100 \, \mathrm{Mpc}$. 

Mildly asymmetric binaries likely represent a more probable and realistic system. In these binaries, both magnetic interaction and electromagnetic emission contribute comparably to the waveform phase, rather than one of the effects dominating. Their horizon distances are expected to be between the two extremes shown for identical field binaries and the highly asymmetric cases discussed here. Thus, the highlighted cases span the range from conservative to optimistic scenarios for binaries with at least one strongly magnetized component. In our analysis, the azimuthal angles are fixed to preserve symmetry, and varying them would lead to small changes in the horizon distances within the band ($e=0-0.4$) for eccentric systems, with no impact on circular binaries.

We illustrate the detectability thresholds for magnetic field effects in a GW170817-like event located at a distance of $ d = 40 \, \mathrm{Mpc}$. If such an event is observed in the future and contains significant magnetic fields, the marked points in~\cref{fig:horizon_distance_a,fig:horizon_distance_b} indicate the minimum field strengths required to leave detectable imprints in gravitational wave data across different observatories. For equal magnetic field configurations, LIGO (O5) can detect fields $ \gtrsim 6 \times 10^{15} \, \mathrm{G} $, ET $ \gtrsim 2 \times 10^{15} \, \mathrm{G} $, and DECIGO $ \gtrsim 5 \times 10^{14} \, \mathrm{G} $. In asymmetric scenarios, where only the primary neutron star is significantly magnetized, these thresholds increase to $ \gtrsim 2 \times 10^{17} \, \mathrm{G} $ for LIGO, $ \gtrsim 5 \times 10^{16} \, \mathrm{G} $ for ET, and $ \gtrsim 10^{16} \, \mathrm{G} $ for DECIGO. A future detection of a GW170817-like event with magnetic fields above these thresholds could provide a unique opportunity to constrain or measure the field strengths in merging neutron star binaries. For reference, we also indicate the distances at which the SNR drops below 8. 

These results emphasize a rare subclass of compact binaries: systems in which one or both neutron stars possess magnetar-level magnetic fields. Although the influence of such fields on gravitational wave signals is extremely subtle, their detection could provide crucial evidence for the existence of these exotic systems. While numerous neutron star binaries have been identified through radio and X-ray observations, binaries containing two magnetars or at least one magnetar have not yet been observed. Gravitational-wave imprints may offer a unique avenue to probe these elusive configurations.

Additionally, gravitational waves also provide a new way to measure magnetic field strengths in neutron star binaries, especially in high-SNR events, where parameter estimation is more precise. The approach is conceptually similar to extracting the tidal deformability, which provides information about a neutron star's compactness and equation of state. In this context, the magnetic field is a direct macroscopic property that, if measured through gravitational wave observations, could advance our understanding of magnetic field evolution in neutron stars, including their decay timescales, maximum limits, and the dynamical role they play in late-stage inspiral and merger dynamics. 

\subsection{Fisher Analysis}\label{section:fisher-matrix}

We employ the Fisher information matrix formalism~\citep{vallisneri2008use, cutler1994gravitational} to estimate how precisely magnetic field strength can be extracted from gravitational wave signals in favorable scenarios. In the high-SNR limit, the posterior over source parameters $\{ \Theta_1, \Theta_2, \dots, \Theta_n \}$ can be approximated by a Gaussian, with a covariance matrix $\Sigma = \Gamma^{-1}$, where the Fisher matrix is defined as: 
\begin{equation}
\Gamma_{ij} = (\partial_i h ,|, \partial_j h),
\label{eq:fisher_matrix}
\end{equation}
with partial derivatives taken with respect to the source parameters $\partial_i \equiv \partial/\partial \Theta_i$~\citep{muttoni2023dark, ma2020excitation}. For a network of $m$ detectors, we have $\Gamma_\mathrm{net} = \sum_{k=1}^{m} \Gamma_k$.  The uncertainty in $\Theta_i$ is given by
\begin{equation}
\Delta \Theta_{i} = \sqrt{\Sigma_{ii}}.
\label{eq:sigma_i}
\end{equation}

As a reference source, we analyze a binary neutron star system similar to GW170817 using the \texttt{GWFish} package~\citep{dupletsa2023gwfish}. The source has component masses of $m_1 = 1.510 \, M_{\odot}$
and $m_2 = 1.254 \, M_{\odot}$, and is located at a luminosity distance of $d_L = 43.74 \, \mathrm{Mpc}$ ~\citep{abbott2017gw170817}. The system is inclined at $\iota = 2.545 \, \mathrm{rad}$ with sky coordinates $(\mathrm{RA},\mathrm{DEC}) = (3.446,-0.408) \,\mathrm{rad}$. The coalescence phase, polarization angle, and coalescence time are fixed to $\phi_c = 0$, $\psi = 0$, and $t_c = 1187008882.4 \, \mathrm{s}$, respectively. A moderate orbital eccentricity of $e = 0.1$ is also assigned at 5 Hz. 

For the waveform, we employ a frequency domain model that incorporates orbital eccentricity and subdominant harmonics, based on postcircular expansion~\citep{yunes2009post,huerta2014accurate}. Magnetic corrections are included in the phase evolution. To isolate the dominant magnetic effect relevant for waveform dephasing, we retain only terms proportional to $\bar{\mathcal{F}}(\vec{\mu}_1, \vec{\mu}_2)$. This captures the magnetic interaction between neutron stars and the coupling to orbital eccentricity. Both neutron stars are assumed to possess equal surface magnetic field strengths of $B_1 = B_2 = B$, with magnetic moments being taken to be aligned along $\hat{L}$. Under these assumptions, only the magnetic field parameter $B$ governs the magnetic influence on the system through: 
\begin{equation}
\bar{\mathcal{F}}(\vec{\mu}_{1}, \vec{\mu}_{2})  =  - \mu_{1} \, \mu_{2} = - B^2 \, \big( \tfrac{1}{2} R_{NS}  \big)^{2}
\end{equation}
In this analysis, the contribution of the effective dipole moment $\vec{\mu}_{\mathrm{eff}}$, associated with electromagnetic radiation, is neglected. For equal field strengths and aligned orientation, this term evaluates to zero and does not affect the parameter estimation. It becomes relevant only in the presence of field asymmetries.

We construct a sample of 20 neutron star binaries with intrinsic and extrinsic parameters resembling GW170817, varying only the magnetic field strength in the range $B \in [10^{15}, 10^{17}] \, \mathrm{G}$. This choice is motivated by the detectability estimates presented earlier: At GW170817-like distances, fields in this range exceed the threshold for detecting magnetic effects. Our focus here is on quantifying the uncertainty in measuring $B$ within this range. The posterior distributions of the source parameters are obtained using Fisher analysis in three detector configurations: the LIGO-Virgo-KAGRA (LVK) network in O5, ET, and the combined ET plus CE network. The SNRs for the simulated events are approximately 140 in the LVK (O5) scenario, 653 in ET, and 2175 in ET+CE \footnote{Since magnetic corrections are incorporated exclusively in the waveform phase, and not in the amplitude, the SNR values are independent of magnetic field strength.}. With such high SNRs, the standard source parameters, including orbital eccentricity, are well constrained and should be reliably recovered. We therefore restrict our discussion to the uncertainty in $B$. 

\begin{figure}
\centering
\includegraphics[scale=0.44]{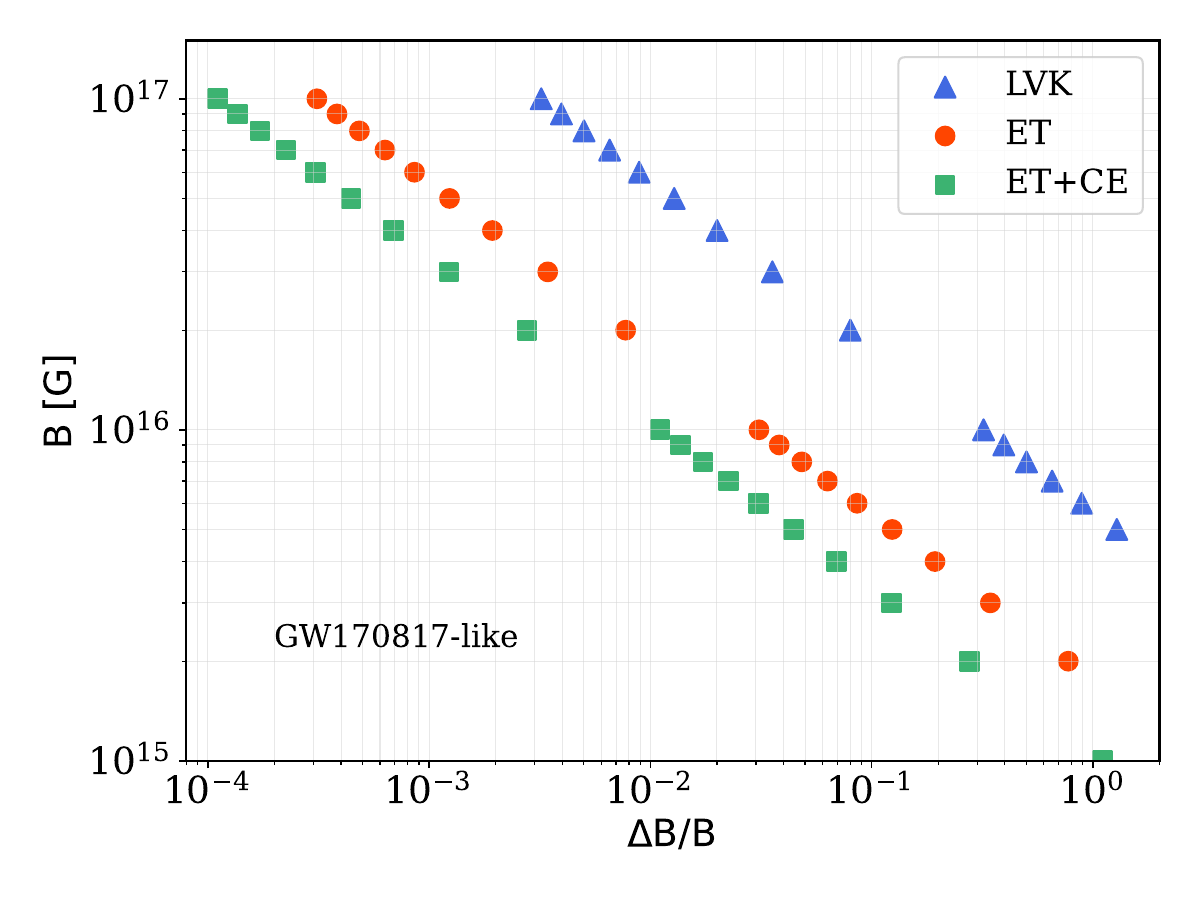}
\caption{Fractional uncertainty in the magnetic field parameter, $\Delta B/B$, estimated from Fisher analysis for a small population of GW170817-like sources at a luminosity distance of $d_{L} = 43.74\,\mathrm{Mpc}$. The magnetic fields of both components are assumed identical and aligned with $\hat{L}$. The system is governed solely by magnetic interaction, with field strengths in the range $10^{15}$–$10^{17}$ G, from magnetar-level to ultrastrong regimes. The measurement uncertainties are large at lower field strengths, but improve to below $10\%$ and in some cases to less than $1\%$, for stronger fields.}
\label{fig:fisher_fractional_uncertainity}
\end{figure}

The fractional uncertainty in the estimation of $B$ is presented in \cref{fig:fisher_fractional_uncertainity}. The large errors (exceeding $100\%$) for some sources stem from their proximity to the detection threshold and suboptimal sky locations. For stronger fields well above the detection threshold, the estimates improve significantly. In ET the uncertainty is $10\%$ for $5 \times 10^{15} \, \mathrm{G}$, and reduces to $1\%$ for $10^{16} \, \mathrm{G}$. For the LVK network, only fields above $ \sim 10^{16} \, \mathrm{G}$ yield uncertainty below $10\%$. Relative to LVK, ET yields lower uncertainties by a factor of 10, with CE providing further improvement. For the strongest fields considered $\sim 10^{17} \, \mathrm{G}$, the ET + CE network achieves a measurement precision of $\Delta B/B \sim 10^{-4}$. The ET+CE network is capable of constraining typical magnetar-level fields ($10^{15}$–$10^{16} \, \mathrm{G}$) with an accuracy of a few percent or better. However, for fields in $10^{15}$–$10^{16} \, \mathrm{G}$ range, LVK is expected to provide only coarse constraints, which are sufficient mainly to exclude extremely strong fields, and performs well in the $10^{16}$-$10^{17} \, \mathrm{G}$ regime, where the accuracy level of a few percent is attainable. These estimates depend on the source distance and will vary for closer or more distant events.

In this analysis, we incorporate the magnetic interaction term in the waveform phase but omit the contribution from the effective dipole moment. Including the net dipole term would likely increase the correlations between the parameters and could lead to larger uncertainties. A more comprehensive analysis framework could include both magnetic terms or prioritize the dominant term depending on which parameter is better measured. Magnetic moments may also be parameterized similarly to spin vectors~\citep{blanchet2024post}, which could assist in inferring their orientations, although this aspect is not explored here. 

On the waveform side, we employed an eccentric waveform model that does not include spin effects and tidal interactions of neutron stars, which may impact parameter uncertainties. Notably, both the leading magnetic interaction and spin-spin coupling terms enter at 2PN order. In binaries where neutron stars possess magnetar-level fields or higher, neutron stars may spin down considerably on short timescales (in $ \sim 10^{2-3}$ years), making these spin-spin terms weaker. The observed magnetars are typically slow-rotating, with characteristic ages of about $10^{3}$ years and a spin frequency of around $\sim 0.1-1 \, \mathrm{Hz}$~\citep{kaspi2017magnetars} (see Fig. 2 of~\cite{kou2019rotational}). The dimensionless spin $\chi$ for such magnetars lies in the range $10^{-3}-10^{-4}$, considerably smaller than the $\chi \sim 0.02$ expected for fast-spinning neutron stars in binary~\citep{bernuzzi2014mergers}. The phase correction due to spin-spin coupling is given by~\citep{blanchet1995gravitational}:
\begin{equation}
\delta \Psi_{ss} = \dfrac{3}{128\, m^{5/3} \eta \, \omega^{5/3}} \left(-10 \, \sigma \, m \,\omega^{4/3} \right),    
\end{equation}
where
\begin{equation}
\sigma = \frac{\eta}{48} \left(-247 \, \chi_{1} \cdot \chi_{2} + 721 \, (\hat{L} \cdot \chi_{1})(\hat{L} \cdot \chi_{2}) \right).    
\end{equation}

For magnetar-like spins, this contribution is roughly two to four orders of magnitude weaker than the phase shift induced by magnetic interaction $\delta \Psi_{\mathrm{mag}}$. Strongly magnetized binaries may therefore be accompanied by low spins, in which case the dominant 2PN correction arises from magnetic effects. In such binaries, magnetic field strength may take precedence over spin in parameter estimation.

It is important to note that in a full Bayesian parameter estimation using real gravitational wave data, uncertainties are expected to exceed those predicted by Fisher matrix estimates for low SNRs \citep{vallisneri2008use}. Since we are primarily interested in high SNR events, this analysis reliably demonstrates the potential for extracting magnetic field information from GW170817-like sources in next-generation detectors.

\begin{figure}
\centering
\includegraphics[scale=0.44]{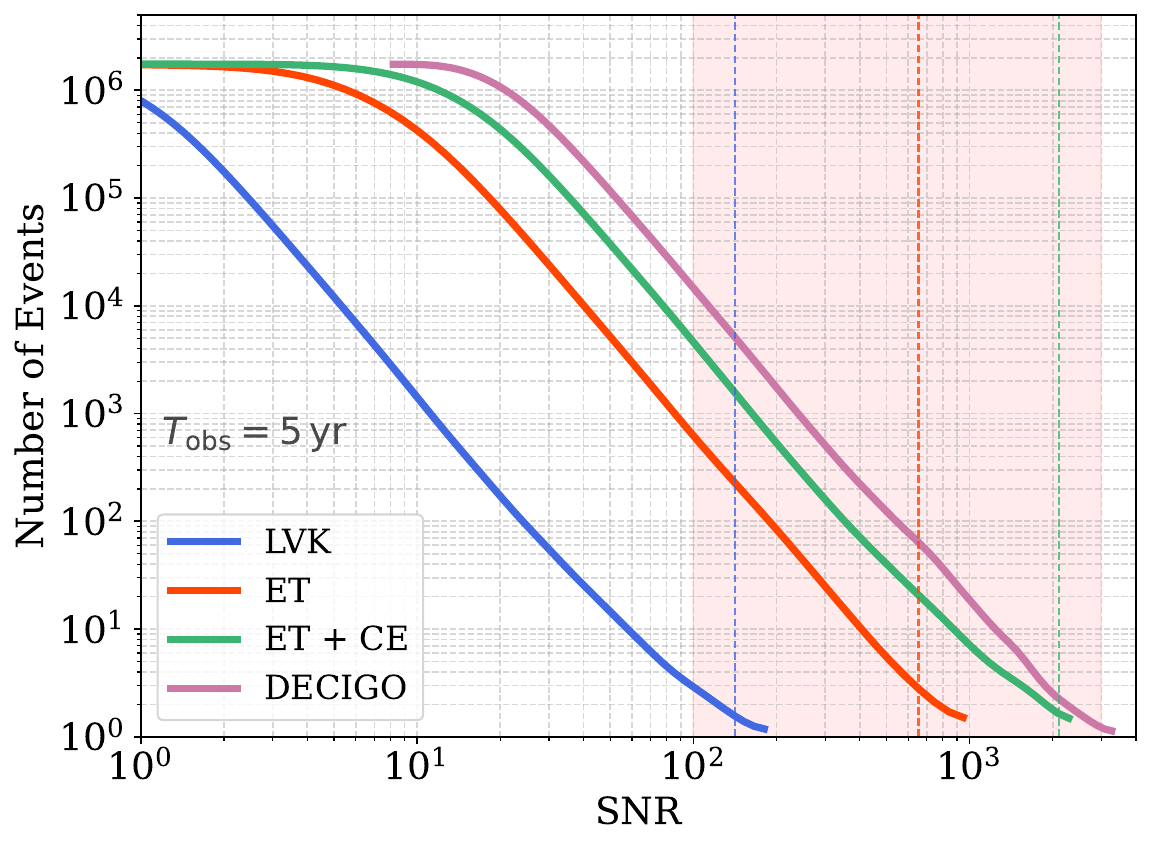}
\caption{Expected number of detected events as a function of SNR for different gravitational wave detector configurations over an observation time of $T_{\mathrm{obs}} = 5$ years. The curves correspond to LIGO-Virgo-KAGRA (LVK) network, ET, ET plus CE network, and DECIGO. The vertical dashed lines indicate the SNR of a GW170817-like event, which is expected to occur roughly once every few years. A substantial number of events are predicted with SNR $>100$ and even $>1000$ for next-generation detectors.}
\label{fig:snr_distribution}
\end{figure}

The detection of magnetic imprints depends on the observation of high-SNR signals. To estimate how frequently magnetic effects can be probed, we generate a population of BNS events using merger rates inspired by star-formation history \citep{huxford2024accuracy,du2025systematic}, along with astrophysical distribution of neutron star masses and binary parameters. For LVK, ET and CE we use the respective antenna patterns, and for DECIGO we use the optimal one. The resulting SNR distribution (Fig.~\ref{fig:snr_distribution}) shows that the number of events with $\text{SNR} \geq 100$ can vary between $10^{2}$ and $10^{4}$, depending on the detector network for an observation time of five years. In addition, $\gtrsim 10$ events reach an SNR of 1000. This abundance underscores the strong potential of future detectors to constrain magnetic field effects in inspiraling binaries.

\section{Conclusion}\label{section:conclusion}

In this work, we explore the influence of strong magnetic fields on the gravitational waves emitted by eccentric neutron star binaries approaching their final inspiral stages. Such systems may arise through dynamical capture involving relatively young neutron stars in dense stellar environments. Although observed neutron-star magnetic fields typically reach $10^{14-15} \, \mathrm{G}$, theoretical considerations permit even stronger fields, potentially up to $\sim10^{17-18} \, \mathrm{G}$. Unlike isolated binaries formed through standard evolutionary channels, where long inspiral times likely lead to significant magnetic-field decay, dynamically assembled binaries in globular clusters or galactic disks may retain their strong magnetic fields until merger. While magnetic contributions are intrinsically weak, their cumulative impact over extended inspirals can become appreciable. Future gravitational-wave detectors with enhanced low-frequency coverage and improved sensitivity may compensate for the weakness of these effects and the subtle imprints they leave on the waveform.

Motivated by these prospects, we analyze the orbital dynamics of magnetized compact binaries within a perturbative framework, treating magnetic field effects as corrections to the leading-order motion. Starting from the Lagrangian of an eccentric binary system with magnetic moments, we analytically derive the orbital solution, the energy loss rates due to gravitational wave emission, and the electromagnetic radiation from the system’s effective dipole, expressed for general magnetic moment orientations. We then obtain the phase evolution of the gravitational wave signal, which consists of contributions from magnetic interaction and electromagnetic dipole radiation. 

We identify two distinct regimes where each effect dominates: symmetric or nearly symmetric configurations ($B_1 \simeq B_2$, or $B_1/B_2 \sim 10^{1–2}$) amplify magnetic interaction effects, while highly asymmetric configurations ($B_1 \gg B_2$) enhance electromagnetic radiation. Magnetic orientation primarily affects the magnetic interaction, with attractive alignments accelerating the inspiral and repulsive ones retarding it. In contrast, effective dipole radiation consistently drives a faster inspiral regardless of orientation. We also uncover coupling terms between eccentricity and magnetic effects that further modulate the signal. Notably, magnetic dephasing scales inversely with frequency, consistent with a 2PN order contribution, whereas tidal dephasing increases with frequency. This qualitative difference may help disentangle magnetic and tidal effects at different stages of the inspiral. 

Assuming identical magnetic fields for both neutron stars, we evaluate the accumulated dephasing governed by magnetic interaction and estimate the optimal SNR required to distinguish magnetic effects in the waveform. Magnetic moments aligned parallel to $\hat{L}$ produce the largest dephasing, perpendicular the least, with intermediate orientations falling in between. Parallel alignment is the most favorable for detectability. Among ground-based detectors, LIGO can identify magnetic fields of $10^{16} \, \mathrm{G}$ at an SNR of 100, whereas ET can probe fields down to $10^{15} \, \mathrm{G}$ at an SNR of 1000. DECIGO's advantage arises primarily from the longer duration over which it can track the signal. It can detect fields as low as few times $10^{14} \, \mathrm{G}$ with SNR of $10^3$. In highly asymmetric configurations, where electromagnetic dipole radiation dominates, stronger fields are required. For such binaries, ET and LIGO can probe magnetic fields of a few $ \times 10^{16} \, \mathrm{G}$ at SNRs of $100-1000$, while DECIGO retains sensitivity to $10^{16} \, \mathrm{G}$ at SNR of $10^3$.

To complement the optimal SNR-based analysis, we evaluate the horizon distance and the minimum magnetic field strength required for distinguishability across different gravitational wave detectors (see Figure \ref{fig:horizon_distance_a} and \ref{fig:horizon_distance_b}). 
This calculation incorporates magnetic corrections in the waveform model and realistic detector sensitivity curves. For binaries with comparable magnetic field strengths, ET and DECIGO can probe fields as low as a few times $10^{13} \, \mathrm{G}$ in nearby galactic systems, and up to $B \sim 10^{16} \, \mathrm{G}$ out to distances of $1$-$10\,\mathrm{Gpc}$. In comparison, LIGO is limited to detecting $B \gtrsim 10^{14} \, \mathrm{G}$ locally and $B \sim 10^{16} \, \mathrm{G}$ out to $\sim 100 \, \mathrm{Mpc}$. For binaries in asymmetric configurations ($B_1 \gg B_2$), DECIGO remains sensitive to $B_{1} = 10^{14} \, \mathrm{G}$ in the local universe and up to $B_{1} \sim 10^{17} \, \mathrm{G}$ at cosmological distances. In contrast, LIGO retains sensitivity to $B_{1} \gtrsim 10^{17} \, \mathrm{G}$ only within a few hundred Mpc. We further examine the parameter estimation performance using Fisher analysis for systems influenced primarily by magnetic interaction. For sources similar to GW170817, the next-generation detector network (ET+CE) can achieve satisfactory measurement of magnetar-level fields ($10^{15}\text{–}10^{16}\, \mathrm{G}$), whereas the LVK (O5) network can provide constraints that are less precise but still informative within this range. Estimates improve significantly for stronger fields, particularly in the range $10^{16}\text{–}10^{17}\, \mathrm{G}$.

These results indicate that current and forthcoming gravitational wave detectors may enable the identification of magnetar binaries, in which one or both neutron stars possess magnetar-level magnetic fields, during their inspiral phase. Such systems are of particular interest, as magnetar-level field systems have not yet been observed through electromagnetic channels. High SNR detections will be crucial to enable such identification. At present, methods for measuring neutron star magnetic fields rely on electromagnetic observations, such as spin-down rates or emission intensities. Gravitational wave observations could offer a novel and complementary approach, potentially enabling direct measurement of strong magnetic fields in compact binaries independent of their electromagnetic visibility. Detecting or ruling out magnetar-level fields in binaries would constrain magnetic field evolution and also shed light on the formation channels of these systems. 

We note that this study does not include spin effects, tidal interactions, or eccentricity terms beyond $\mathcal{O}(e^2)$. This choice allows us to isolate and highlight the role of distinct magnetic effects: magnetic interactions, electromagnetic emissions, and their coupling with orbital eccentricity and magnetic orientations. Although higher-order eccentricity terms become relevant at large eccentricities, they do not qualitatively alter our conclusions. We plan to incorporate these aspects into future work for a more comprehensive analysis. In an alternative approach, some of the authors also explore a parameterized framework that includes magnetic effects \citep{ghosh2025generalized}. 

Strong magnetic fields may also induce significant shape deformations in neutron stars, potentially enhancing waveform modulation. Magnetic fields can also lead to orbital precession, offering another distinctive signature. These effects are not explored in the present study but may provide further avenues for identifying magnetized binaries through gravitational wave observations.

\section{Acknowledgements}
The authors thank Parameswaran Ajith, Rajes Ghosh, and members of the Astrophysics and Relativity group at ICTS for valuable discussions and input. The authors also thank Paolo Pani and Kenta Kiuchi for their helpful comments. Part of this work was carried out during visits to the Institute for Computational and Experimental Research in Mathematics (ICERM) at Brown University and the Albert Einstein Institute in Potsdam; RP gratefully acknowledges their financial support. We also acknowledge support from the Department of Atomic Energy, Government of India, under project number RTI4001. P.K. also acknowledges support from the Ashok and Gita Vaish Early Career Faculty Fellowship at the International Center for Theoretical Sciences. \\

\textit{Software}: \texttt{NumPy} \citep{van2011numpy}, \texttt{SciPy}, \citep{virtanen2020scipy}, \texttt{Matplotlib} \citep{hunter2007matplotlib}

\appendix

\section{Maximum Magnetic Field Supported by a Neutron Star}\label{app:max-mag-field-virial}
We estimate the maximum magnetic field a neutron star can support by equating magnetic energy to the gravitational binding energy, as constrained by the virial theorem. Assuming a uniform magnetic field $ B$ within a neutron star of radius $R$, the total magnetic energy is given by
\begin{equation}
E_B = \frac{B^2}{8\pi} \frac{4}{3} \pi R^3 = \frac{B^2 R^3}{6}.
\end{equation}
We treat the neutron star as a constant-density sphere. For such a star of mass $M$ and radius $R$, the gravitational binding energy is 
\begin{equation}
E_G = -\frac{3}{5} \frac{G M^2}{R}.
\end{equation}
Requiring that magnetic energy does not exceed gravitational binding energy, we obtain
\begin{equation}
B_{\text{max}}  \lesssim \sqrt{ \frac{18}{5}  \frac{G M^2}{R^4} }.
\end{equation}
Substituting the mass and radius values of a very compact neutron star, $M = 2 \, M_\odot$ and $R = 10 \, \text{km}$, gives 
\begin{align}
B_{\text{max}} & \lesssim \sqrt{ \frac{18}{5} \cdot \frac{6.674 \times 10^{-8} \cdot (3.98 \times 10^{33})^2 }{(1.0 \times 10^6)^4} }, \nonumber \\
\implies B_{\text{max}} & \lesssim  1.9 \times 10^{18} \, \mathrm{G}.
\end{align}
This gives an upper bound on the magnetic field that a neutron star can support without being disrupted by magnetic pressure. In real stars, where density and magnetization vary with radius, this upper limit may shift slightly. Rotation can also affect this limit, typically reducing it by a few percent. The general relativistic solutions of rotating magnetized stars place this upper limit at about $\sim 10^{18}$ G~\citep{bocquet1995rotating}.

\section{Orbit-Averaging Relations}\label{app:orbit-averaging}

In a binary neutron star system, the unit separation vector is given by $\hat n = \vec{r}/r$. It lies in the orbital plane and points along the relative separation between the stars, and varies with the angular coordinate as $\hat n = (\cos\phi \, \hat x + \sin\phi  \, \hat y)$. The projections of neutron star magnetic moments along $\hat n$ are given by
\begin{align}
\vec{\mu}_{1} \cdot \hat n &= (\mu_{1x} \hat x + \mu_{1y} \hat y + \mu_{1z} \hat z) \cdot  (\cos\phi \hat x + \sin\phi \hat y), \nonumber \\
\vec{\mu}_{2} \cdot \hat n &= (\mu_{2x} \hat x + \mu_{2y} \hat y + \mu_{2z} \hat z)  \cdot  (\cos\phi \hat x + \sin\phi \hat y). 
\end{align} 
The product can be written as
\begin{align}
(\vec{\mu}_{1} \cdot \hat n) (\vec{\mu}_{2} \cdot \hat n) & =   \mu_{1x} \, \mu_{2x}  \cos^{2}\phi  +  \mu_{1x} \, \mu_{2y}\,  \cos\phi  \, \sin\phi \nonumber  \\  & + \mu_{1y} \, \mu_{2x}\,  \cos\phi  \, \sin\phi  + \mu_{1y} \, \mu_{2y}  \,\sin^{2} \phi .   \nonumber
\end{align}
We perform an orbital average of the above quantity over $\phi$, using $ \langle \mathrm{X} \rangle = \tfrac{1}{2 \pi} (1 - e^2)^{3/2} \int_{0}^{2 \pi} \mathrm{X}(\phi) \, (1 + e \cos\phi)^{-2} \, d\phi$. By averaging and expanding to $e^{2}$, we obtain
\begin{align}
& \left\langle \, (\vec{\mu}_{1} \cdot \hat{n})(\vec{\mu}_{2} \cdot \hat{n}) \, \right\rangle \nonumber \\
& = \frac{1}{2} \, \left( \mu_{1x} \,\mu_{2x} + \mu_{1y} \, \mu_{2y} \right) 
+  \frac{3}{8} \, e^{2} \left( \mu_{1x} \, \mu_{2x} - \mu_{1y} \,\mu_{2y} \right) \nonumber \\
&= \frac{1}{2} \, [ (\vec{\mu}_{1} \cdot \vec{\mu}_{2}) - (\vec{\mu}_{1} \cdot \hat{L})(\vec{\mu}_{2} \cdot \hat{L}) ] \nonumber \\
&  + \, \frac{3}{8} \, e^{2} \left[ (\vec{\mu}_{1} \cdot \hat{p})(\vec{\mu}_{2} \cdot \hat{p}) - (\vec{\mu}_{1} \cdot \hat{q})(\vec{\mu}_{2} \cdot \hat{q}) \right] 
\end{align}
where $\hat{L}$ is a unit vector along the direction of the orbital angular momentum. $\hat{p}$ points towards the periastron, and $\hat{q}$ orthogonal to it in the orbital plane. The first term represents the relative orientations of the magnetic moments, and the second term captures the eccentricity-dependent anisotropy in the orbital plane. For configurations such as identical components in the orbital plane, the anisotropic correction vanishes. For such cases, we can write
\begin{equation}
\left\langle \, \vec{\mu}_{1} \cdot \hat{n})(\vec{\mu}_{2} \cdot \hat{n}) \, \right \rangle
\simeq \frac{1}{2} \left[(\vec{\mu}_{1} \cdot \vec{\mu}_{2}) - (\vec{\mu}_{1} \cdot \hat{L})(\vec{\mu}_{2} \cdot \hat{L})\right].
\end{equation}

Eccentricity corrections enter separately through the perturbative expansion of $r(\phi)$ and the orbit-averaged energy loss as well. The same expression is also obtained in~\cite{ioka2000gravitational}. 

\section{Energy Loss Rate Due to the Motion of an Effective Dipole}\label{app:em-radiation}

We begin with the general expression for the instantaneous energy loss rate due to the motion of a magnetic dipole obtained by~\cite{ioka2000gravitational}:
\begin{align}\label{eq:energy-radiated-em}
\left(\frac{dE}{dt}\right)_{\text{EM}} = &-\frac{2}{15} \frac{m^2}{r^6} \big[ 
    2\mu_{\mathrm{eff}}^2 
    \left( v^2 - 6\dot{r}(\hat{n} \cdot \vec{v}) + 9\dot{r}^2 \right) \nonumber \\
&\quad  - \left\{ \mu_{\mathrm{eff}} \cdot \left( \vec{v} - 3\dot{r} \hat{n} \right) \right\}^2 \, \big],
\end{align}

where $\vec{\mu}_{\mathrm{eff}} = \frac{1}{m}(m_2 \vec{\mu}_1 - m_1 \vec{\mu}_2)$ is the effective dipole moment in terms of neutron star masses and magnetic moments. We assume that the binary motion is confined to a orbital plane. The velocity vector is expressed as $\vec{v} = \dot{r} \hat n + r \dot{\phi} \hat\phi$, with $\hat{n} = (\cos\phi, \sin\phi, 0)$ and $\hat{\phi} = (-\sin \phi,\cos\phi, 0)$. For the radial coordinate, we use an eccentric orbit solution with magnetic interaction corrections, as obtained in~\cref{eq:r_with_deltar}. 
For convenience, we also define an in-plane basis $(\hat{x}, \hat{y})$, with $\hat{x}$ aligned along the periastron and $\hat{y}$ orthogonal to it. We substitute $\vec{v}$ and $\vec{\mu}_{\mathrm{eff}} = (\mu_x, \mu_y, \mu_z)$ in Eq.~\eqref{eq:energy-radiated-em}, and after averaging over an orbit, we get
\begin{align}\label{eq:energy-radiated-em-cartesian}
\left\langle \frac{dE}{dt} \right\rangle = & - \frac{1}{60} M^{2/3} \omega^{14/3} \big[ 12 \mu_x^2 + 12 \mu_y^2 + 16 \mu_z^2 \nonumber \\
& \quad  +  e^2 \left(179 \mu_x^2 + 181 \mu_y^2 + 240 \mu_z^2 \right) \big].
\end{align}
To express the effective dipole moment in terms of orientation angles, we define $\vec{\mu}_{\mathrm{eff}}$ using angles $(\alpha, \beta)$:
\begin{equation}
\vec{\mu}_{\mathrm{eff}} = \mu_{\mathrm{eff}} \left( \sin\alpha \cos\beta, \; \sin\alpha \sin\beta, \; \cos\alpha \right),
\end{equation}
where $\alpha$ is the angle $\vec{\mu}_{\mathrm{eff}}$ makes with orbital angular momentum $\hat{L}$, and $\beta$ is the azimuthal angle in the orbital plane. Substituting the squared components of $\vec{\mu}_{\mathrm{eff}}$ in Eq.~\eqref{eq:energy-radiated-em-cartesian}, we obtain

\begin{align}
\left\langle \frac{dE}{dt} \right\rangle &= 
- \frac{1}{15} m^{2/3} \omega^{14/3} \mu_{\mathrm{eff}}^2 \big[
3 \sin^2\alpha + 4 \cos^2\alpha  \nonumber \\
& + e^2 \, (45 \sin^2\alpha + 60 \cos^2\alpha - \tfrac{1}{4} \cos2\beta \sin^2\alpha) \big]. \nonumber
\end{align}
This result can be expressed compactly in terms of angular functions as
\begin{align}
\left\langle \frac{dE}{dt} \right\rangle & =
- \frac{1}{15} m^{2/3} \omega^{14/3} \mu_{\mathrm{eff}}^2 
\left[ \mathcal{F}_0(\alpha) + e^2 \, \mathcal{F}_1(\alpha, \beta) \right],  \nonumber \\[1ex]
\mathcal{F}_0(\alpha) &= 3 \sin^2\alpha + 4 \cos^2\alpha, \nonumber \\[1ex]
\mathcal{F}_1(\alpha, \beta) & = 45 \sin^2\alpha + 60 \cos^2\alpha - \tfrac{1}{4} \cos2\beta\sin^2\alpha.
\end{align}
The magnitude of $\vec{\mu}_{\mathrm{eff}}$ and its orientation angles $(\alpha,\beta)$ govern the energy loss in electromagnetic radiation. For $\vec{\mu}_{\mathrm{eff}}$ with symmetric in-plane components, that is, $\mu_{\mathrm{eff},x} = \mu_{\mathrm{eff},y}$, the dependence on $\beta$ drops out, and $\mathcal{F}_1(\alpha, \beta)$ simplifies to $\mathcal{F}_1(\alpha) \approx 45 \sin^2\alpha + 60 \cos^2\alpha$. Also, for scenarios that involve varying or random orientations in the plane over an orbit, averaging over $\beta \in [0, 2\pi]$ gives $ \left\langle \cos2\beta \right\rangle = 0$.

\subsection{In-Plane Case: $\vec{\mu}_{\mathrm{eff}} \perp \hat{L}$}

We set $\alpha = \pi/2$, which implies $\cos\alpha = 0$ and $\sin\alpha = 1 $. In this configuration, the effective dipole lies entirely on the orbital plane:
\begin{equation}
\vec{\mu}_{\mathrm{eff}} = \mu_{\mathrm{eff}} (\cos\beta, \sin\beta, 0).
\end{equation}
Substituting into the expression of the energy loss rate gives
\begin{equation}
\left\langle \frac{dE}{dt} \right\rangle =
- \frac{1}{15} m^{2/3} \omega^{14/3} \mu_{\mathrm{eff}}^2 \big[ 3 + e^2 \big(45 - \frac{1}{4} \cos2\beta  \big) \big]. \nonumber
\end{equation}
If $ \vec{\mu}_{\mathrm{eff}}$ has equal components in the plane $\mu_{\mathrm{eff,x}} = \mu_{\mathrm{eff,y}}$. Under this azimuthal symmetry of the effective dipole moment in the orbital plane, the $\beta$ dependence vanishes. The expression then simplifies to
\begin{equation}
\left\langle \frac{dE}{dt} \right\rangle = -\frac{1}{5} m^{2/3} \omega^{14/3} \mu_{\mathrm{eff}}^2 \left[ 1 + 15 e^2 \right],
\end{equation}
which matches the circular orbit result obtained by~\cite{ioka2000gravitational} in the $e=0$ limit, and extends to eccentric binaries with azimuthally symmetric dipole orientations. 

\subsection{Perpendicular-to-Plane Case: $\vec{\mu}_{\mathrm{eff}} \parallel \hat{L}$}

Here, we set $\alpha = 0$, which implies $\cos\alpha = 1$ and $\sin\alpha = 0$. For this orientation, the dependence of $\beta$ vanishes. The effective dipole is aligned with the orbital angular momentum:
\begin{equation}
\vec{\mu}_{\mathrm{eff}} = \mu_{\mathrm{eff}} (0, 0, 1).
\end{equation}
Substituting into the orbit-averaged energy loss rate equation gives
\begin{equation}
\left\langle \frac{dE}{dt} \right\rangle =
- \frac{4}{15} m^{2/3} \omega^{14/3} \mu_{\mathrm{eff}}^2 \left[ 1 + 15 e^2 \right],
\end{equation}
which also reproduces the circular orbit result obtained by~\cite{ioka2000gravitational} in the $e = 0$ limit.

\bibliography{main}{}
\bibliographystyle{aasjournal}

\end{document}